\documentclass[preprint2]{proto}
\usepackage{natbib}      
\bibpunct{(}{)}{;}{a}{,}{,}   
\usepackage{times}        

\usepackage{graphicx}

\newcommand{\teff}{\mbox{T$_{eff}$}}
\newcommand{\mjup}{\mbox{M$_{\rm Jup}$ }}
\newcommand{\mearth}{\mbox{M$_\oplus$}}
\newcommand{\msun}{\mbox{M$_\odot$}}

\begin{document}

\title{\textbf{\LARGE Giant planet and brown dwarf formation}}

\author {\textbf{\large Gilles Chabrier}}
\affil{\small\em Ecole Normale Sup\'erieure de Lyon\\University of Exeter}
\author {\textbf{\large Anders Johansen}}
\affil{\small\em  Lund University}
\author {\textbf{\large Markus Janson}}
\affil{\small\em Princeton University\\Queen's University Belfast}
\author {\textbf{\large Roman Rafikov}}
\affil{\small\em Princeton University}

\begin{abstract}

 Understanding the dominant brown dwarf and giant planet formation processes, and finding out whether these processes rely on completely different mechanisms or share common channels represents one of the major challenges of astronomy and remains the subject of heated debates. 
 It is the aim of this review to summarize the latest developments in this field and to address the issue of origin by confronting different brown dwarf and giant planet formation scenarios to presently available observational constraints. 
 As examined in the review, if objects are classified as "Brown Dwarfs" or "Giant Planets" on the basis of their formation mechanism,  it has now become clear that their mass domains overlap and
that there is no mass limit between these two distinct populations.
  Furthermore, while there is increasing observational evidence for the existence of non-deuterium burning brown dwarfs, some giant planets, characterized by a significantly metal enriched composition, might be massive enough to ignite deuterium burning in their core. Deuterium burning
(or lack of) thus plays no role in either brown dwarf or giant planet formation. Consequently, we argue that the IAU definition
to distinguish these two populations has no physical justification and brings scientific confusion.
In contrast, brown dwarfs and giant planets might bear some imprints of their  formation  mechanism, notably in their mean density and in the
physical properties of their atmosphere. Future direct imaging surveys will undoubtedly provide crucial information and perhaps provide 
some clear observational diagnostics to unambiguously distinguish these different astrophysical objects.

\baselineskip = 11pt
\leftskip = 0.65in 
\rightskip = 0.65in
\parindent=1pc
{\small .\\~\\~\\~}
     
\end{abstract}

\section{\textbf{INTRODUCTION}}  

The past decade has seen a wealth of sub-stellar object (SSO) discoveries. SSO's are defined as objects not massive enough to sustain hydrogen burning and thus encompass brown dwarfs (BD) and planets, more specifically giant planets (GP), we will focus on in the present review. 
For sake of clarity, we will first specify which objects we refer to as "brown dwarfs" and "giant planets", respectively.
For reasons which will be detailed in the review, we do {\it not} use the IAU definition to distinguish between these two populations. Instead, we adopt a classification which, as argued along the lines below, (1) appears to be more consistent with observations, (2) is based on a physically motivated justification, invoking two drastically different {\it dominant} formation mechanisms, (3) allows a mass overlap between these two populations, as indeed supported by observations but excluded by the present official definition. This classification is the following. We will denominate "brown dwarfs" (1) all {\it free floating objects} below the hydrogen-burning minimum mass, $\sim 0.075\,\msun$ for solar composition \citep{Chabrier00a}, no matter whether or not they are massive enough to ignite deuterium burning, and (2) objects that are companions to a parent star or another BD, but exhibit compositional and mechanical (mass-radius) properties consistent with the ones of a gaseous sphere of {\it global chemical composition similar to the one of the parent star/BD}. In contrast, "giant planets" are defined as 
companions of a central, significantly more massive object, a {\it necessary condition}, whose bulk properties strongly depart from the ones just mentioned. We are well aware of the fact that these definitions do retain some ambiguity in some cases (e.g. some planets might be ejected onto an hyperbolic orbit and become genuine "free floating planets") but, as argued in the review, these ambiguous cases are likely to be statistically rare. 
We will come back to this issue in the conclusion of the review.

According to the above classification,
the detection of BD's now extends down to objects nearly as cool as our jovian planets, both in the field and in cluster associations, closely approaching the bottom of the stellar initial mass fonction (IMF). 
Extrasolar planets, on the other hand, are discovered by radial velocity techniques and transit surveys at an amazing pace. The wealth of discoveries now extends from gaseous giants of several Jupiter masses
 to Earth-mass objects. 
 
Both types of  objects  raise  their  own  fundamental  questions  and
generate their own debates. Do brown dwarfs form  like  stars,  as  a
natural extension of the stellar formation process in the substellar  regime? Or are brown  dwarfs  issued  from  a  different  formation  scenario  and
represent a distinctive class  of  astrophysical  bodies ?  Similarly,  can
giant planets form in a fashion  similar  to  stars,  as  the  result  of  a
gravitational instability in a protoplanetary disk, or do they  emerge  from
a  completely  different  process,  involving   the   early   accretion   of
planetesimals? What are the distinctive observational signatures  of  these
different processes?  From  a  broader  point  of  view,  major  questions
concerning the very identification  of  brown  dwarfs  and  gaseous  planets
remain to be answered. Among which, non exhaustively:  do the  mass domains
of brown dwarfs and giant planets overlap or is there a clear  mass  cut-off
between these two populations?  Does  deuterium  burning  (resp.  lack  of)
play a determinant role in brown dwarf (resp. planet)  formation,  providing
a key diagnostic to distinguish these two populations  or  is  this  process
inconsequential? With a connected, crucial question : what is  the  minimum
mass for  star/brown  dwarf  formation  and  the  maximum  mass  for  planet
formation? 

     It is the aim of the present review to address these questions and  to
find out  whether  or  not  we  have  at least some reasonably  robust  answers about BD and GP dominant formation mechanisms
by confronting predictions of the various  suggested
formation scenarios with observations.  
The  general  outline  of the review is as follows.
In sections 2 and 3, we first summarize  the   most   recent   observational
determinations of brown dwarf and giant planet properties. This includes  both
field and star forming regions for brown  dwarfs  as  well  as  transit  and
radial velocity surveys for planets. We will particularly  focus  on  direct
imaging observations of Jupiter-mass bodies as wide companions of a  central
star, which may encompass both types of objects. In \S4, we briefly summarize the observable
properties of protostellar and protoplanetary disks and examine the required conditions for such disks 
to be gravitational unstable and fragment. Section 5 is devoted to a brief summary of
the different existing formation scenarios for  brown
dwarfs and planets. We will identify the distinctive concepts  and  physical
mechanisms at the heart  of  these  scenarios  and  examine  their  viability by confronting them with  the  previous  observational
constraints. In section 6, we will examine whether BD's and GP's can be distinguished according to various signatures which have been suggested in the literature, namely early evolution, deuterium burning or mass domain.
Finally, as the  outcome  of  these  confrontations,  we  will  try in the conclusion to
determine what is or are the {\it dominant} formation  mechanisms  for
brown dwarfs and planets and  whether  or not they may share some common channels.

\section{\textbf{OBSERVATIONAL PROPERTIES OF BROWN DWARFS}} 
\label{IMF}

\begin{figure*} 
\epsscale{1.5}
\plotone{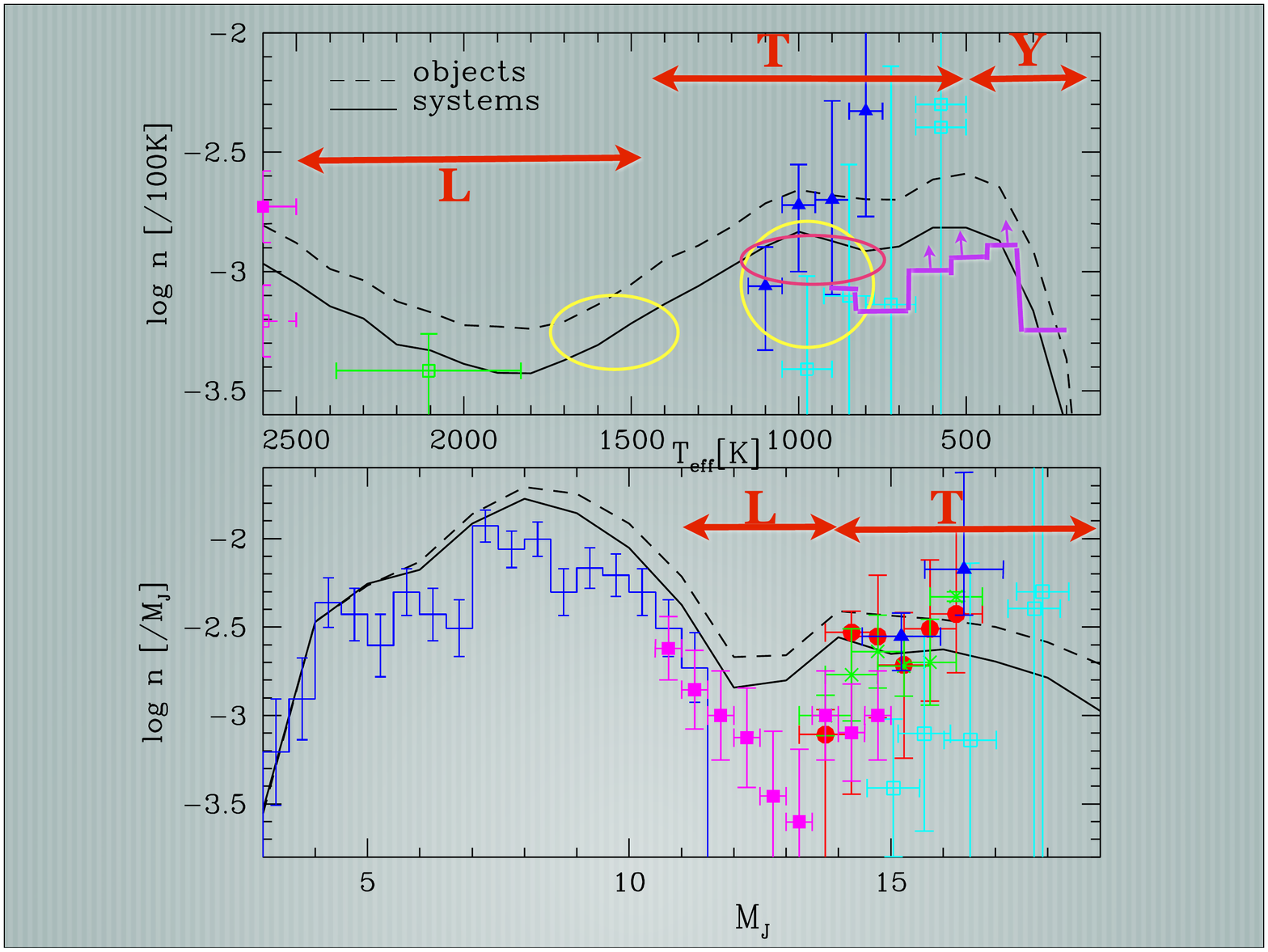}
\caption{\small Brown dwarf number density as a functiion of $\teff$ (top) and absolute J-magnitude (bottom) predicted with a Chabrier (2005) IMF for resolved objects (dash) and unresolved binary systems (solid). The L, T and Y letters correspond to the various BD spectral types. The solid triangles, squares and circles indicate various observational determinations \citep{Burgasser04, Cruz07, Metchev08, Burningham10, Reyle10}, while the histogram with the arrows portrays the WISE (uncomplete) observations \citep{Kirkpatrick11}.
}  
\label{fig_imf}
\end{figure*}

An excellent summary of BD observational properties can be found in \cite{Luhman07} and \cite{Luhman12}. We refer the reader to these reviews, and   all appropriate references therein, for a detailed examination of BD observational properties. We will only briefly summarize some results here. 

\subsection{\textbf{The brown dwarf census and mass function}} 
\label{BD_IMF}

It is now widely admitted 
that the stellar initial mass function, defined hereafter as $dN/d\log M$ 
\citep{Salpeter55}, flattens below a mass $\sim 0.6 \,\msun$ compared to a Salpeter IMF \citep{Kroupa01, Chabrier03}.
Although the IMF below this mass is often parametrised by a series of multiple power-law segments, such a form does not have physical foundation. In contrast, the IMF can also be parametrised by a power-law+lognormal form, a behavior which has been shown to emerge from the gravoturbulent picture of star formation \citep{HC08,Hopkins12}. The behavior of such a stellar IMF in the BD regime can now be confronted directly to observations  both in the field and in young clusters and star-forming regions.
Figure 1 compares the predictions of the number density distributions of BD's as a function of $\teff$ and J-absolute magnitude for an age of $\sim$1 Gyr, assuming a Galactic constant star formation rate, a double-exponential Galactic disk density distribution and a Chabrier (2005; hereafter C05) IMF (see \cite{Chabrier02,Chabrier05} for details of the calculations), with the most recent observational determinations. The C05 IMF is based on a slighly revised version of the Chabrier (2003) IMF \citep{Chabrier03} (see {\it Chabrier}, 2005). The observations include in particular the recent results from the 
WISE survey, which nearly reaches the bottom end of the BD mass distribution ($\teff<500$ K) in the field \citep{Kirkpatrick11}. It
is striking to see how the WISE discoveries, as well as the ones from previous surveys, are consistent with a C05 IMF extending from the stellar regime into to the BD one. As shown in Table 1 of Chabrier (2005), the BD-to-star ratio predicted with this IMF was $\sim$1/4, while the ratio revealed so far by the (still partly incomplete at the $\sim$ 20\% level) WISE survey is $\sim$1/5.  
The determination of the age of the WISE objects is an extremely delicate task. Evolutionary models (mass, age) and synthetic spectra ($\teff$, $\log\,g$) in this temperature range remain so far too uncertain to provide a reliable
answer. The WISE survey, however, probes the local thin-disk population, so the discovered BDs should be on average less than about 3 Gyr old. According to BD theoretical evolution models \citep{Baraffe03, burrows2003, SaumonMarley08}, for an age of 3 Gyr (resp. 1 Gyr) BD's with effective temperatures below 400 K (resp. $\sim$550-600 K) must be less massive than 0.012 $\msun$, i.e. below the D-burning limit for solar composition  \citep{Chabrier00b}. 
So far, the WISE survey has even discovered 1 field BD with $\teff \lesssim 350$ K, reaching the temperature domain of Jupiter-like objects. As just mentioned, the determination of the effective temperature in this regime still retains significant uncertainties and these values should be taken with due caution but these results suggest that some of the field objects discovered by WISE lie below the D-burning limit. At any rate, the WISE survey, as both statistics and models improve, holds the perspective to nail down the minimum mass for star/BD formation and the IMF, a result of major importance.

Notice that a C05 IMF would correspond globally to a negative $\alpha$ value, in terms of $dN/dM\propto M^{-\alpha}$, below about $\teff\lesssim 1300$ K, indicating a {\it decreasing number of BD's with decreasing mass below $\sim 30\,M_{\rm Jup}$} for a age of 1 Gyr. This is in stark contrast 
with the {\it planetary} mass function, which reveals an {\it increasing number of objects} below about the same mass \citep{Mayor11, howard2010}.
Similarly, the BD mass distribution has now been determined in more than a dozen clusters or star-forming regions down to $\sim 0.02\,\msun$ or less and these distributions are found to agree quite well, with some expected scatter, with the very same C05 IMF for unresolved systems (see e.g. Fig. 8 and 9 of \cite{Jeffries12}). 

Brown dwarfs are now routinely discovered with masses below the D-burning limit
\citep{Lucas06,Caballero07,LuhmanMuench08,Marsh10,Scholz12a,Scholz12b,Lodieu11,Delorme12}. Although these objects are sometimes denominated  free floating "planetary mass objects", based on the IAU definition, such a distinct appelation is not justified, as these objects are regular genuine BDs not massive enough to ignite D-burning. The same way stars not massive enough to ignite H-burning via the CNO cycles are still called "stars" and have never been named differently.

All these BD discoveries are consistent with the {\it very same underlying "universal" IMF} extending continuously from the stellar regime into the BD one down to about the opacity fragmentation limit, with no noticeable variations between various regions (under Milky Way like conditions) nor evidence whatsoever for a discontinuity near the H-burning limit. 
This strongly argues in favor of a scenario in which BD's and stars share dominantly the same formation mechanism. 

There has been some claims based on microlensing observations \citep{Sumi11} for an excess of a very-low mass object population ($\sim$ a few $\mjup$) in the field,
formed in proto-planetary disks and subsequently scattered into unbound orbits. The validity of these results,
however, can be questioned for several reasons. First of all, the excess number is based on 5 events - which depart form the expected distribution only at the $\sim$1$\sigma$ limit - and thus has no statistical significance. Second, when observing a short duration
{\it isolated event} with low S/N and non-complete
coverage, the symmetry of the event, a characteristic property of lensing events, is very 
difficult to assess properly, in contrast to the case when the planetary event
occurs as a caustic during a stellar lensing event. Free-floating planet candidate events are thus much more uncertain than star-planet candidate events and should be considered with extreme caution.
Third, planet-planet scattering usually produces objects with high velocity dispersions compared with the parent star. As examined below (\S\ref{BD_vel}), no such population of high velocity objects in the BD or planet mass range has been discovered so far, at least down to present observational limits.
Most importantly, the excess of low-mass objects claimed by the microlensing results would correspond to a
population of free-floating Jupiter-mass objects in the field almost twice as common as stars, a census clearly excluded so far by observations \citep[e.g.][]{Scholz12a}.

\subsection{\textbf{Brown dwarf velocity and spatial distributions}} 
\label{BD_vel}

Available radial velocity measurements of BD's in  young clusters and  star-forming regions clearly indicate that young BD's and young stars have similar velocity dispersions \citep[e.g.][]{Joergens06, WhiteBasri03, Kurosawa06}. Similar observations show that BD's have the same spatial
distributions as stars (e.g. \cite{Parker11,Bayo11,Scholz12a} for recent determinations). These observations clearly contradict at least the
early versions of the accretion-ejection model for BD formation (\ref{sec:acc_eje}) \citep{ReipurthClarke01, KroupaBouvier03} which predicted that young BD's should have higher velocities than stars and thus should be more widely distributed. 


\subsection{\textbf{Brown dwarfs or giant planets as star or brown dwarf companions}} 
\label{BD_cn}

The statistics of brown dwarf and giant planet companions to stars or BD's potentialy bears important information on the formation process of these objects. Various studies have indeed 
compared the multiplicity distributions of prestellar core fragmentation
simulations with the presently observed values in young clusters or star
forming environments. 
Such comparisons, however, are of limited utility to infer the very formation process. Indeed, while simulations address the very prestellar cloud/core collapse stages, i.e. the very initial epoch ($\sim 10^5$ yr) of star formation, observations probe present conditions, i.e. typical ages of 1 to several Myr's. Dynamical interactions and interactions with
 the disk have kept occuring between these two epochs and have 
strongly influenced the multiplicity properties. It is well establihed and intuitively expected, for instance, that the most weakly
bound systems, which include preferentially those with BD or very-low-mass star companions, are more prone to disruption than more massive ones, so that low-mass binary properties are expected to differ substantially from their primordial distribution \citep[e.g.][]{DucheneKraus13}. This naturally explains the decreasing binarity frequency with mass (see below). Therefore, presently observed values are of limited relevance to infer primordial multiplicity properties and their dependence on formation conditions. 

Keeping that in mind, observations today point to star and BD binary properties which appear to be continuous from the stellar to the
BD regime, with binary fractions and average separations decreasing continuously with primary mass, and mass ratios increasing with it \cite[e.g.][]{Faherty11,KrausHillenbrand12,Luhman12}. The continuity of the multiplicity properties
thus mirrors the one found for the IMF. 

An important observational constraint for BD formation theories also stems from the observations of 
wide binaries. A particular example is the wide binary FU Tau a,b, with no identified stellar companion in its vicinity \citep{Luhman09}. 
This is the best example that BDs can form in isolation, with no nearby stellar companion, in low-density environments (stellar density $n_\star \sim 1$-10 pc$^{-3}$). 
Several wide ($\approx 400$-4000 AU) binary substellar companions, down to $\sim 5\,\mjup$\citep{Luhman11,Aller13}, have also been observed in star-forming regions.
One of these objects has an extreme mass ratio $q=M_2/M_1\approx 0.005$. Worth mentioning are also the isolated and rather fragile BD wide binary 2MASS J12583501/J12583798 (6700 AU) \citep{Radigan09} or the quadrupole system containing 2MASS 04414489+23015113 (1700 AU) \citep{Todorov10}.
All these systems seem rather difficult to
explain with dynamical processes or disk fragmentation and clearly indicate that  these mechanisms are not {\it necessary} for BD formation. As discussed in \S\ref{bin_fn}, such wide binaries are most likely the
indication that core fragmentation into binaries might occur at the very early stages of star formation (see \S\ref{bin_fn} below).

Other relevant observational inputs for understanding BD and GP formation include the brown dwarf desert \citep[e.g.][]{grether2006}, which is a lack of companions in the mass range 10--100 $M_{\rm Jup}$ (i.e., 1--10\% of the primary mass) in the separation range covered with radial velocity (up to a few AU) around Sun-like stars. The frequency of companions increases with smaller mass below the desert \citep[e.g.][]{howard2010}, and increases with larger mass above the desert, clearly indicating separate formation mechanisms between planetary and stellar companions. Moreover, giant planets appear to become less frequent around lower-mass stars and more frequent around higher-mass stars (e.g. \citealt{johnson2010}), and meanwhile, brown dwarf companions become more frequent around low-mass stars and brown dwarfs in the same separation range \citep[e.g.][]{joergens2008}. This further implies that the desert is primarily determined by system mass ratios, and is not specific to brown dwarf secondaries. 
As mentioned in \S\ref{BD_IMF}, the fact that the mass function of planetary companions to stars \citep{Mayor11, howard2010} differs drastically, {\it qualitatively and quantitatively}, from the BD mass function, clearly points to a different dominant formation mechanism for these two populations.
More quantitative results about the statistics of wide orbit SSO's around stars and BD's in the BD/GP mass range will be given in \S\ref{sec:DI}.

\subsection{\textbf{Observational identifications of isolated proto- and pre-brown dwarf cores}} 
\label{BDcore}

The recent results of the HERSCHEL survey confirm that the mass function of pre-stellar cores (the so-called CMF) in molecular clouds resembles the stellar IMF down to the present observational sensitivity limit, which lies within the BD domain \citep{Andre10, Konyves10} (see also chapter by {\it Offner et al.}). Although caution is necessary at this stage before definitive answers can be claimed, these results strongly argue in favor of the stellar+BD IMF been determined at the very early stages of star formation 
rather than at the final gas-to-star conversion stage, an issue we will come back to in \S\ref{sect:formation_BD}.

Several class 0 objects have been identified with {\it Spitzer} surveys with luminosities $L<0.1\,L_\odot$, indicative of young proto-BDs \citep[e.g.][]{Huard06,Dunham08, Kauffmann11} (see further references in {\it Luhman}, 2012). Although some of these objects may eventually accrete enough material to become stars, some have such a small accretion reservoir that they will definitely remain in the BD domain \citep{LeeKim09, Kauffmann11, Lee13}. 
Most importantly,
there are now emerging discoveries of {\it isolated proto-brown dwarfs} \citep{Lee09b, Lee09a, Palau12, Lee13}. The existence of such isolated proto-BD's again indicates that dynamical interactions, disk fragmentation or photoionizing radiation are not essential for BD formation.

Of noticeable importance also is the discovery of the isolated, very low-mass brown dwarf OTS 44 (spectal type M9.5, $M\sim12\, \mjup$), with significant accretion and a substantial disk \citep{Joergens2013}. This
demonstrates that the processes that characterize the canonical star-like mode of formation apply to isolated BD's
down to few Jupiter masses. 

But the most striking result concerning BD formation is the recent observation of an isolated {\it pre}-brown dwarf core \citep{Andre12}. The mass and size of this core have been determined with the PdBI, yielding a core mass $\sim 30\,\mjup$ with a radius $<460$ AU, i.e. a mean density $n\sim 10^7$-$10^8$ cm$^{-3}$, consistent with theoretical predictions of the gravo-turbulent scenario (\S\ref{sec:gravo}). It should be stressed that this core lies {\it outside} dense filaments of Oph and thus is not expected to accrete a significant amount of mass (see Supplementary Material of \cite{Andre12}). Indeed, the surrounding gas has been detected in follow-up observations with the IRAM telescope and the most conservative estimates suggest at most a factor 2 growth in mass, leaving the final object in the BD regime. This is the unambiguous observation of the possibility for BD cores to form in isolation by the collapse of a surrounding mass reservoir, as expected from gravoturbulent cloud fragmentation.
Given the nature of the CMF/IMF, however, the number of low-mass cores is expected to decrease exponentially, making their discoveries very challenging.

\subsection{\textbf{Summary}}
\label{BD_sum}

In summary (again see \cite{Luhman12} for a more throughout review), it seems rather robust to affirm that young brown dwarfs and young stars: (1) exhibit similar radial velocity dispersions; (2) have similar spatial distribution in young clusters; (3) have mass distributions consistent with the same underlying IMF; (4) have both wide binary companions; (5) exhibit similar accretion + disk signatures such as large blue/UV excess, with large asymetric emission lines (UV, H$_\alpha$, Ca II etc...),
with accretion rates scaling roughly with the square of the mass, ${\dot M}\propto M^2$, from 2 down to about $\sim 0.01\,\msun$, signatures all consistent with a natural extension of the CTT phase in the BD domain; (6) have similar disk fractions; (7) both exhibit outflows; (8)  have both been identified at the early pre-stellar/BD stage of core formation in isolated environments. It seems that the most natural conclusion we can draw at this stage from such observational results is that 
star and BD formations seem to share many observational signatures, pointing to a common {\it dominant} underlying mechanism, and that
dynamical interactions, disk fragmentation or photoionizing radiation are not essential for BD formation.

\section{\textbf{OBSERVATIONAL PROPERTIES OF GIANT PLANETS}} 

\subsection{\textbf{Observed correlations}} 
\label{sec:correl}


During the two decades over which exoplanetary observation has been an active area in astronomy, the field has expanded rapidly, both in terms of the number of detected planets and in terms of the parameter space of planets that has opened up for study. As a result, giant planets can now be studied with good completeness over a rather wide domain of parameters. By combining radial velocity and transits, both radius and mass as well as a range of other properties can be studied, and additional constraints can be acquired from astrometry and microlensing, as well as direct imaging, which is described separately below. Furthermore, the host star properties can be studied and correlated with planet occurrence, in order to gain additional clues about planet formation.

One of the most well-known and important relations between host star properties and planet occurrence is the planet-metallicity correlation \citep[e.g.][]{santos2004,fischer2005}, which refers to the rapid increase in giant planet frequency as a function of stellar host metallicity. While only a few percent of solar abundance stars have giant planets with orbital periods of less than 4 years, $\sim$25\% of stars with with [Fe/H] $>$0.3 have such planets. This is a strong piece of evidence in favour of core accretion {(hereafter CA) as a formation scenario, since a framework in which planets grow bottom-up from solids in a protoplanetary disk favors a high concentration of dust and ices. Equally interesting in this regard is the fact that the correlation appears to weaken for lower planetary masses \citep[e.g.][]{sousa2008,Mayor11,buchhave2012}. This again fits well into the core accretion scenario, since it would be natural to assume that the gas giant planets are most dependent on high concentrations of solids, as they need to rapidly reach the critical core mass for gas accretion, before the gas disk dissipates. It also appears that planet frequency and/or mass scales with stellar mass \citep[e.g.][]{lovis2007,hekker2008,johnson2010}, although there is not as clear a convergence on this issue in the literature as for the metallicity correlation \citep[see e.g.][]{lloyd2011,mortier2013}. In any case, this is once again easy to understand in the core accretion paradigm, as more mass in the star means more mass in the disk, and thus a more generous reservoir from which to grow planets rapidly and to large sizes.

An aspect in the study of giant planets that has received considerable attention in recent years is the study of alignment (or mis-alignment) between the orbital plane of the planet and the rotational plane of the star. A projection of the differential angle between these planes can be measured in transiting systems using the Rossiter-McLaughlin (RM) effect \citep{rossiter1924,mclaughlin1924}. The RM effect is a sequence of apparent red/blue-shift of the light from a rotating star, as a planet blocks approaching/receding parts of the stellar surface during transit. This is the most commonly used method for measuring spin-orbit alignment \citep[e.g.][]{queloz2000,winn2005}, but other methods have recently been implemented as well. These include spot crossings during transit \citep[e.g.][]{sanchis2012} and independent determinations of the inclination of the stellar plane; the latter can be accomplished through relating the projected rotational velocity to the rotational period and radius of the star \citep[e.g.][]{hirano2012}, or through observing mode splitting in asteroseismology \citep[e.g.][]{chaplin2013}. 

In almost all of the above cases, the measurement only gives a projection of the relative orientation. As a consequence, a planet can be misaligned, but still appear aligned through the projection, so it is typically not possible to draw conclusions from single cases. However, firm statistical conclusions can be drawn from studying populations of objects. The studies that have been performed show that giant planets are sometimes well-aligned with the stellar spin, but sometimes strongly misaligned or even retrograde \citep[e.g.][]{triaud2010}. Since there are indications for a correlation between orbital misalignment and expected tidal timescale, where systems with short timescales have low obliquities \citep{albrecht2012}, it has been hypothesized that all systems with a close-in giant planet may feature a misalignment between the stellar spin and planetary orbit, but that alignment occurs over tidal timescales. There is a possible contrast here with lower-mass planets, which often occur in multiply transiting configurations \citep[e.g.][]{Lissauer11b,johansen2012,fabrycky2012}. This implies a high degree of co-planarity, at least between the planes of the different planets. Furthermore, the few stellar spin measurements that have been performed for small-planet systems imply co-planarity also in a spin-orbit sense \citep[e.g.][]{albrecht2013}. If these trends hold up with increasing data, it could indicate that while gas giants and smaller planets probably pass through a similar stage of evolution (core formation), there may be differences in the evolutionary paths that bring them into small orbits.

Since transiting giant planets can be relatively easily followed up with radial velocity, both mass and radius can be determined in general for such planets, which allows to make inferences about their composition. Due to nearly equal compensation between electron degeneracy and interionic electrostatic interactions \citep{CB00}, an isolated gas giant planet without a substantial core or other heavy element enrichment will have a radius that only very weakly depends on mass. Indeed, the radius will consistently be close to that of Jupiter, apart from during the earliest evolutionary phases, when the planet is still contracting after formation \citep[e.g.][]{Baraffe03,burrows2003}. However, with a larger fraction of heavy elements, the radius decreases, thus providing a hint of this compositional difference \citep{Fortney07, Baraffe08, Leconte09}. In practice, when studying transiting planets for this purpose, there is a primary complicating factor: the planets under study are far from being isolated, but rather at a small distance from their parent star. The influence of the star may substantially inflate the planet, sometimes bringing the radius as high as 2~$R_{\rm Jup}$ \citep[e.g.][]{hartman2011,anderson2011,Leconte09}. Many different mechanisms have been proposed to account for precisely how this happens, 
but it remains so far an unsolved and interesting problem. 
For the purpose of our discussion, however, we merely establish that it is a complicating factor when
 evaluating the composition of the planet. Nonetheless, some transiting gas giants do feature sub-Jovian radii (e.g. HD~149026~b, \citealt{sato2005}; HAT-P-2b, \citealt{bakos2007}), which has led to the inference of very large cores with tens or even hundreds of Earth masses \citep[e.g. $>$200 $\mearth$ for HAT-P-2b;][]{Baraffe08,Leconte09,Leconte11}. Although it is unclear how exactly these large cores could form, such a heavy material enrichment can hardly be explained by accretion of solids on a gaseous planet embryo, as produced by gravitational instability (see \S\ref{sect:GI_physics}). This may thus imply that the formation of these objects occurred through the
process of core accretion, possibly involving collisions between Jupiter-mass planets (\citealt{Baraffe08}).

Other relevant observational inputs for understanding giant planet formation include the brown dwarf desert and the brown dwarf mass function, an issue we have already addressed in \S\ref{BD_cn}. The observations show that
 the frequency of companions increases with smaller mass below the desert and increases with larger mass above the desert \citep{Mayor11, howard2010}, clearly indicating separate formation mechanisms between planetary and stellar/BD companions.
 
Finally, both microlensing data \citep[e.g.][]{gould2010} as well as the asymptotic trends in radial velocity data \citep[e.g.][]{cumming2008} indicate that giant planets are more common beyond the ice line than they are at smaller separations. This may again constitute a clue regarding the formation, although it should probably be regarded primarily as providing constraints on migration, rather than formation.

\subsection{\textbf{Direct imaging constraints}} 
\label{sec:DI}

Thanks to developments in both adaptive optics-based instrumentation and high-contrast imaging techniques, direct imaging of wide massive planets has become possible over the past decade. This has led to a number of surveys being performed, ranging from large-scale surveys, covering a broad range of targets \citep[e.g.][]{lowrance2005,chauvin2010}, to more targeted surveys of specific categories, such as low-mass stars \citep[e.g.][]{masciadri2005,delorme2012}, intermediate-mass stars \citep[e.g.][]{lafreniere2007,biller2007,kasper2007}, high-mass stars \citep[e.g.][]{janson2011,vigan2012,nielsen2013}, or stars hosting debris disks \citep[e.g.][]{apai2008,janson2013,wahhaj2013}. A common trait of all such surveys is that they generally aim for quite young targets (a few Myr to at most a few Gyr). This is due to the fact that distant giant planets (which are not strongly irradiated by the primary star) continuously cool over time, and hence become continuously fainter with age, which increases the brightness contrast between the planet and star and makes the planet more difficult to detect at old stellar ages. 

The number of giant exoplanets that have been imaged so far is problematic to quantify exactly. This is partly due to the lack of a clear consensus concerning the definition of what a planet is, in terms of (e.g.) mass versus formation criteria, and partly due to uncertainties in evaluating a given object against a given criterion. For instance, the errors in the mass estimate often lead to an adopted mass range that encompasses both deuterium-burning (DB) and non-DB masses. Likewise, it is often difficult to assess the formation scenario for individual objects, given the relatively few observables that are typically available. As a result, the directly imaged planet count could range between $\sim$5 and $\sim$30 planets in a maximally conservative and maximally generous sense, respectively. Two systems that would arguably count as planetary systems under any circumstance are HR~8799 \citep{marois2008,marois2010} and $\beta$~Pic \citep{lagrange2009,lagrange2010}. All the four planets around HR~8799 and the single detected planet around $\beta$~Pic have best-fit masses well into the non-DB regime ($\sim$5--10~$M_{\rm Jup}$), mass ratios to their primaries of less than 1\%, and indications of sharing a common orbital geometry, both with each other (for the HR~8799 system) and with the debris disks that exist in both systems. Recently, another planet has been imaged around the Sun-like star GJ~504 \citep{kuzuhara2013}, which has a similarly small mass ratio and a mass that remains in the non-DB regime independently of the choice of initial conditions in existing theoretical mass-luminosity relationships, and thus a similar reliability in its classification.
The most puzzling properties of the HR~8799 planets in a traditional planet-forming framework are their large orbital separations ($\sim$10--70~AU for HR~8799~b/c/d/e), as will be further discussed in section \ref{sect:formation_GP}. 

Besides the above relatively clear-cut cases, it is informative to probe some of the more ambiguous cases which may or may not be counted as planets, depending on definitions and uncertainties. The B9-type star $\kappa$~And hosts a companion with a mass of $\sim$10--20~$M_{\rm Jup}$ \citep{carson2013}, straddling the DB limit. 
However, due to the very massive primary, the mass ratio of the $\kappa$~And system is only 0.5--1.0\%. This is very similar to the HR~8799 and $\beta$~Pic planets, and the companion resides at a comparable orbital separation of $\sim$40~AU. This might imply a similar formation mechanism, which would support the fact that planet formation can produce companions that approach or exceed the DB limit, as has been suggested by theoretical models (see chapter by {\it Baraffe et al.}). 

A candidate planet presently in formation has been reported in the LkCa15 system \citep{kraus2012}. 
With the potential for constraining the timescale of formation and
initial conditions of a forming planet, such systems would have a clear
impact on our understanding of planet formation, but more data will be
needed to better establish the nature of this candidate before any
conclusion can be drawn in that regard.
The young Solar analog 1RXS~1609 has a companion with a mass of $\sim$6--11~$M_{\rm Jup}$ \citep{lafreniere2008} and a system mass ratio of ~1\%, but the projected separation is 330~AU, which may be larger than the size of the primordial disk. It is therefore unclear if it could have formed in such a disk, unless it was, e.g., scattered outward in three-body interactions. One of the clearest hints of incompatibility between a DB-based and a formation-based definition is 2M~1207~b \citep{chauvin2005}, which is a $\sim$4~$M_{\rm Jup}$ companion to a $\sim$20~$M_{\rm Jup}$ brown dwarf \citep[e.g.][]{mamajek2005}. Thus, the companion is firmly in the non-DB mass regime, but with a system mass ratio of 20\%, the couple appears to be best described as an extension of the brown dwarf binary population, rather than a planetary system. 

Despite these detections, the general conclusion from the direct imaging surveys performed so far is that very wide and massive objects in the giant planet mass domain are rare, with many of the surveys yielding null results. These results can be used to place formal constraints on the underlying planet population. For example, the GDPS survey \citep{lafreniere2007} has indicated that the fraction of FGKM-type stars harboring $>$2~$M_{\rm Jup}$ objects in the 25--420~AU range is less than 23\%, at 95\% confidence. For $>$5~$M_{\rm Jup}$ objects, the corresponding upper limit is 9\%. In order to examine how this translates into total giant planet frequencies, \citet{lafreniere2007} assume a mass distribution of $dN/dM \propto M^{-1.2}$ and a semi-major axis distribution of $dN/da \propto a^{-1}$. This leads to the conclusion (at 95\% confidence) that less than 28\% of the stars host at least one 0.5--13~$M_{\rm Jup}$ planet in the semi-major axis range of 10--25~AU, less than 13\% host a corresponding planet at 25--50~AU, and less than 9\% at 50--250~AU. \citet{nielsen2010} combine the GDPS data with spectral differential imaging (SDI) data from VLT and MMT. They find that if power law distributions in mass and semi-major axis that fit the radial velocity distribution \citep{cumming2008} are adopted, and extrapolated to larger distances, planets following these particular distributions cannot exist beyond 65~AU at 95\% confidence. This limit is moved out to 82~AU if stellar host mass is taken into consideration, since the most stringent detection limits in imaging are acquired around the lowest-mass primaries, which is also where the giant planet frequency is expected to be the lowest from RV data \citep[e.g.][]{johnson2010}.

The above studies assume a continuous distribution from small separation
planets to large separation planets, and thus essentially assume the same
formation mechanism (primarily core accretion) for all planets. However, at
wide separations, disk instability becomes increasingly efficient (and core
accretion less efficient), such that one might imagine that two separate
populations of planets exist, formed by separate mechanisms. Hence, an
alternative approach to statistically examine the direct imaging data is to
interpret them in the context of predictions from disk instability formation.
This was done in \citet{janson2011}, where the detection limits from direct
imaging were compared to the calculated parameter range in the primordial disk
in which a forming planet or brown dwarf could simultaneously fulfill the
Toomre criterion and the cooling criterion for fragmentation (see \S4.2 for a discussion of these criteria)}, for a sample of
BA-type stars. It was concluded that less than 30\% of such stars can host disk
instability-formed companions at 99\% confidence. This
work was later extended to FGKM type stars in \citet{janson2012} using the GDPS sample, where
additional effects such as disk migration were taken into account, assuming a
wide range of migration rates from virtually no migration to very rapid
migration. There, it was found that {\it $<$10\% of the stars could host disk
instability-formed companions at 99\% confidence}. Since a much larger fraction
than this are known to host planets believed to have formed by core accretion
\citep[e.g.][]{Mayor11,batalha2013}, it {\it strongly suggests that disk
instability cannot be a dominant mechanism of planet formation}. This
upper limit is tightened further by the discovery of rapid branches of the core
accretion scenario which can lead to formation of gas giants in wide orbits, as
discussed below in \S\ref{sect:formation_GP}.

It is important to note that several assumptions are included in the various analyses described above, some of which are quite uncertain. In particular, all of the above studies translate brightness limits into mass limits using evolutionary models that are based on so-called "hot-start" conditions \citep[e.g.][]{Baraffe03,burrows2003} (see \S\ref{hot_cold}). It has been suggested that the initial conditions may be substantially colder \citep[e.g.][]{fortney2008} or have a range of values in between \citep{spiegel2012}, which could make planets of a given mass and age significantly fainter than under hot-start conditions. This would lead to overly restrictive upper limits on planet frequency in the statistical analyses. However, the measured temperatures of most of the planets that have been imaged so far are consistent with hot start predictions, and robustly exclude at least the coldest range of initial conditions \citep[e.g.][]{marleau2013,Bowler13}, an issue we will address 
further in  \S\ref{hot_cold}. As will be shown in this section, cold vs hot/warm-start conditions are not necessarily equivalent to formation by core accretion vs disk instability.
Nonetheless, the fact remains that the detection limits are based on mass-luminosity models that have not been fully calibrated in this mass and age range, which is an uncertainty that should be kept in mind. In any case, direct imaging surveys have already started to put meaningful constraints on the distributions of distant giant planet mass objects, and near-future instruments such as SPHERE \citep{beuzit2008}, GPI \citep{macintosh2008} and CHARIS \citep{peters2012} will substantially enhance the capacity of this technique, thus leading to tighter constraints and more detections of closer-in and less massive objects. 

Aside from providing information about the distributions of planet mass objects, the future instruments mentioned above will also be able to provide substantial information about atmospheric properties, through their capacity of providing planetary spectra. This has already been achieved in the HR~8799 system with present-day instrumentation \citep[e.g.][]{janson2010,bowler2010,oppenheimer2013,konopacky2013}, but can be delivered at much higher quality and for a broader range of targets with improved instruments. Spectroscopy enables studies of chemical abundances in the planetary atmospheres, which may in turn provide clues to their formation. For instance, core accretion might result in a general metallicity enhancement, similar to how the giant planets in the Solar System are enriched in heavy elements with respect to the Sun. This may be possible to probe with metal-dominated compounds such as CO$_2$ \citep[e.g.][]{lodders2002}. Another possibility is that the formation may leave traces in specific abundance ratios, such as the ratio of carbon to oxygen \citep[e.g.][]{madhusudhan2011,oberg2011}. The C/O ratio of planets should remain close to stellar if they form through gravitational instability. However, if they form by core accretion, the different condensation temperatures of water and carbon-bearing compounds may lead to a non-stellar abundance ratio in the gas atmosphere that is accreted onto the planet during formation, unless it gets reset through subsequent planetesimal accretion. The most robust observational study addressing this issue to date is that of \citet{konopacky2013}, but while the results do indicate a modestly non-stellar C/O ratio, they are not yet unambiguous. Again, future studies have great promise for clarifying these issues.

\section{PROTOSTELLAR AND PROTOPLANETARY DISKS}
\label{sect:disk}

\subsection{Observations of protostellar and protoplanetary disks}
\label{sect:obs_disks}

Circumstellar disks around young objects are generally studied via their thermal and line emission at IR and mm wavelengths, where they dominate the stellar radiation. Unfortunately, the determination of the radii of most disks cannot be done with current technology. Noticeable exceptions are nearly edge-on disks which can be detected in scattered light while occulting the star. Using {\it Spitzer} spectroscopy and high-resolution imaging from {\it HST}, Luhman et al. (2007b) were able to
determine the disk radius of a young BD in Taurus, and this latter was found to be small size, with $R\approx 20$-40 AU. 

This is also what seems to be found at the early class 0 stage with observations of isolated, massive, compact ($R\sim 20$ AU) disks \citep{Rodriguez05} 
while so far large and massive disks prone to fragmentation appear to be very rare at this stage of evolution \citep{Maury10}. 
Two noticeable exceptions are the $R\sim 70$-120 AU, $M\sim 0.007\;\msun$ disk around the $\sim 0.2\;\msun$ protostar L1527 \citep{Tobin13a} and the $R\sim 150$ AU, $M\sim 0.02\;\msun$ disk around VLA1623 \citep{Murillo13}. 
Moreover, \cite{Tobin13b} observed density structures around the Class 0/I protostars CB230 IRS1 and L1165-SMM1 that they identify as companion protostars, because of the strong ionizing radiation, along with an extended ($\sim$ 100 AU), possibly circumbinary disc-like structure for L1165-SMM1. 
Whether these companions indeed formed in an originally massive disk or during the very collapse of the prestellar core itself, however, remains an open question.
In contrast, the extended emission around the Class 0 L1157-mm is suggestive of a rather small ($\lesssim$ 15 AU) disc.
It should be noticed, however, that L1527, CB230 IRS1 and L1165-SMM1 are more evolved than genuine Class 0 objects like L1157-mm, and
 protostellar discs are expected to expand significantly with time due to internal angular momentum redistribution \citep[e.g.][]{Basu98,Machida10,Dapp12}.
 Large disks are indeed observed at later Class 0/I or Class I-II stages (see chapter by Z.-Y. Li et al.). At such epochs, however, they are unlikely to be massive enough relative to the central star to be prone to fragmentation (see next section). Indeed, observed Class I disks have generally masses less than 10\% of the star mass.
 
Clearly, this issue of massive discs at the early stages of star formation is not quite settled for now and will strongly benefit from ALMA.
In any case, as discussed in detail \S\ref{sect:formation_disk}, compact disk observations are in agreement with MHD simulations and thus suggest that magnetic field is a key element in 
shaping up protostellar disks.


\subsection{Disk instability}
\label{sect:GI_physics}

The idea that massive gaseous disks may fragment under the action of their own gravity into bound, self-gravitating objects is rather old \citep{Kuiper51}. However, the quantitative analysis of its operation was not around until the works of \cite{Safronov60} and \cite{GLB65}. They were the first to show that a 2D (i.e. zero thickness) disk of rotating fluid becomes unstable due to its own gravity provided that the so-called Toomre $Q$ satisfies
\begin{eqnarray}
Q\equiv\frac{\Omega c_s}{\pi G\Sigma} < 1,
\label{eq:Q}
\end{eqnarray}
where $c_s$ and $\Sigma$ are the sound speed and surface density of the disk. This particular form assumes a Keplerian rotation, when the so-called epicyclic frequency coincides with the angular frequency $\Omega$.

The physics behind this criterion is relatively simple and lies in the
competition between the self-gravity of gas, which tends to destabilize the
disk, and stabilizing influence of both pressure and Coriolis force. If we
consider a parcel of gas with characteristic dimension $L$ and surface density
$\Sigma$ in a zero-thickness differentially rotating disk, one can evaluate the
attractive force it feels due to its own self-gravity as $F_G\sim
G\Sigma^2L^2$. However, at the same time it feels a repulsive pressure force
$F_P\sim c_s^2\Sigma L$, where $c_s$ is the sound speed. In addition,
differential rotation of the disk causes fluid element to spin around itself at
the rate $\sim \Omega$, resulting in a stabilizing centrifugal force $F_C\sim
\Omega^2 \Sigma L^3$. Gravity overcomes pressure ($F_G>F_P$) when the gas
element is sufficiently large, $L\gtrsim c_s^2/(G\Sigma)$, which is in line
with the conventional Jeans criterion for instability in a three-dimensional
homogeneous self-gravitating gas. However, in a disk case self-gravity also has
to exceed the Coriolis force ($F_G>F_C$), which results in an upper limit on the
fluid parcel size, $L\lesssim G\Sigma/\Omega^2$. Gravity wins, and fluid
element collapses when both of these inequalities are fulfilled, which is only
possible when the condition (\ref{eq:Q}) is fulfilled. This condition is often
recast in an alternative form, as a lower limit on the local volumetric gas
density $\rho\gtrsim M_\star/a^3$ at semi-major axis $a$.

Applying this criterion to a disk around Solar type star one finds that disk gravitational instability (hereafter GI) can occur only if the disk surface density is very high,
\begin{eqnarray}
\Sigma> \frac{\Omega c_s}{\pi G}\approx 10^5\mbox{g cm}^{-2}\left(\frac{\mbox{AU}}{r}\right)^{7/4},
\label{eq:Sig_lim}
\end{eqnarray}
where we used temperature profile $T(r)=300\,\mbox{K}\,(r/\mbox{AU})^{-1/2}$. This
considerably (by more than an order of magnitude) exceeds surface density in
the commonly adopted minimum mass solar nebula (MMSN). Such a disk extending to $50$ AU would have a mass
close to $M_\odot$. For that reason, with the development of the core accretion
paradigm in the end of 1970th the interest in planet formation by GI has gone
down, but see \citet{Cameron78}.

Situation started changing in the 1990s, when problems with the CA scenario for
planet formation at large separations mentioned in \S2 started emerging.
Stimulated by this, \citet{Boss97} resurrected the GI scenario in a series of
papers, which were based on grid-based numerical simulations of
three-dimensional disks. These calculations adopted isothermal equation of
state (the validity of this assumption will be discussed below) and found that massive disks with $\Sigma$ exceeding the limit
(\ref{eq:Sig_lim}) indeed readily fragment, which would be expected if disk
fragmentation was governed by the criterion (\ref{eq:Q}) alone.

However, the logic behind linking the outcome of gravitational instability
unambiguously to the value of Toomre $Q$ is not without caveats. Indeed, both
the simple qualitative derivation presented above and the more rigorous
analyses of the dispersion relation explicitly assume that pressure of
collapsing fragment is the unperturbed pressure. This is a good approximation
in the initial, linear phase of the instability, meaning that Toomre $Q$ is a
good predictor of whether the instability can be triggered. However, in the
nonlinear collapse phase gas density grows and pressure increases, eventually
stopping contraction provided that $F_P$ grows faster than $F_G$ as the gas
element contracts. In real disks this is possible e.g. if contraction is
isentropic and gas has adiabatic index $\gamma>4/3$, which is the case for
molecular gas in protoplanetary disks. As a result, in the absence of radiative
losses gravitational instability cannot result in disk fragmentation, which
agrees with numerical calculations.

Cooling of collapsing fluid elements lowers their entropy thus reducing pressure support and making fragmentation possible. Using 2D hydro simulations in the shearing-sheet approximation \citet{Gammie01} has shown that fragmentation of a gravitationally unstable disk patch is possible only if the local cooling time of the disk $t_c$ satisfies
\begin{eqnarray}
t_c\Omega\lesssim \beta,
\label{eq:cool_crit}
\end{eqnarray}
where $\beta\approx 3$. In the opposite limit of long cooling time the disk settles into a quasi-stationary gravitoturbulent state, in which it is {\it marginally} gravitationally unstable.

A simple way to understand this result is to note that the fluid element contracts and rebounces on dynamical timescale $\sim \Omega^{-1}$, so unless cooling proceeds faster than that, pressure forces would at some point become strong enough to prevent collapse. This result has been generally confirmed in different settings, e.g. in global 3D geometry, and with both grid-based and SPH numerical schemes \citep{Rice05}. The only difference found by different groups is the precise value of $\beta$ at which fragmentation becomes possible, for example \citet{Rice05} find $\beta\approx 5$ in their 3D SPH calculations. The value of $\beta$ is also a function of the gas equation of state (EOS), with lower $\gamma$ resulting in higher $\beta$ \citep{Rice05}. This is natural as gas with softer EOS needs less cooling for the role of pressure to be suppressed. It should nevertheless be noted that some authors found the cooling criterion (\ref{eq:cool_crit}) to be non-universal, with $\beta$ dependening on resolution of simulations, their duration and size of the simulation domain \citep{Meru11,Paardekooper12}. It is not clear at the moment whether these claimed deviations from simple cooling criterion have physical or numerical origin \citep{Paardekooper11,Meru12}.

Cooling time in a disk can be estimated using a simple formula
\begin{eqnarray}
t_c\sim \frac{\Sigma c_s^2}{\sigma T_{eff}^4}=\frac{\Sigma (k_B/\mu)}{\sigma T^3f(\tau)},
\label{eq:t_cool}
\end{eqnarray}
where $T_{eff}$ is the effective temperature at the disk photosphere, related to the midplane temperature $T$ via $T^4=T_{eff}^4 f(\tau)$.
The explicit form of the function $f(\tau)$ depends on a particular mode of vertical energy transport in the disk, but it is expected to scale with the optical depth of the disk $\tau$. As disk cooling is inefficient both in the optically thick ($\tau\gg 1$) and thin ($\tau\ll 1$) regimes, $f(\tau)\gg 1$ in both limits. In particular, if energy is transported by radiation then $f(\tau)\approx \tau+\tau^{-1}$ and similar expression has been derived by \citet{Rafikov07} if energy transport is accomplished by convection. Interestingly, 
the cooling constraint is easier to satisfy at
higher disk temperatures since, for a fixed $f(\tau)$, the radiative flux from the disk surface increases with temperature as $T^4$. 
This is a much faster scaling than that of the thermal content of the
disk, which is linear in $T$. Thus, to satisfy equation (\ref{eq:t_cool}) one would like to increase $T$, forcing $\Sigma$ to grow to keep $Q\lesssim 1$. As a result, at a given radius, fragmentation of a gravitationally
unstable disk is possible only if the disk is hot enough  (short cooling time) and very dense \citep{Rafikov05, Rafikov07}.

Even under the most optimistic assumptions about the efficiency of cooling ($f(\tau)\sim 1$) a gravitationally unstable disk with a surface density profile (\ref{eq:Sig_lim}) has
\begin{eqnarray}
t_c\Omega\sim \frac{\Omega^2}{\pi G\sigma}\left(\frac{k_B}{\mu}\right)^{3/2}T^{-5/2}\approx 600\left(\frac{\mbox{AU}}{r}\right)^{7/4}
\label{eq:GI_cool}
\end{eqnarray}
for the previously adopted temperature profile. Fragmentation criterion (\ref{eq:cool_crit}) is clearly not fulfilled at separations of several AU in such a disk, making planet or BD formation via GI problematic in the inner regions of the protoplanetary disk \citep{Rafikov05}. \citet{Boss02} resorted to convective cooling as a possible explanation of rapid heat transport between the disk midplane and photosphere but \citet{Rafikov07} has demonstrated that this is unlikely possibility. Subsequently, \citet{Boley06} have interpreted vertical motion seen in \citet{Boss02} simulations as being the shock bores rather than the true convective motions.

Combining gravitational instability and cooling criteria (\ref{eq:Q})
and (\ref{eq:t_cool}) one can make more rigorous statements and show that disk  fragmentation is very difficult to achieve close to the star, within several tens of AU \citep{Rafikov05,Rafikov07}. In particular, for the disk opacity scaling as $\kappa\propto T^2$ thought to be appropriate for outer, cool disk regions, fragmentation of the optically thick disk is inevitable outside $60-100$ AU around a Solar type star \citep{Matzner05,Rafikov09}, but cannot occur interior to this zone. This result may be relevant for understanding the origin of the outermost planets in the HR8799 system, although, as discussed in \S\ref{sect:formation_GP}, these planets might also have formed by core accretion.

In optically thin disks this boundary is pushed considerably 
further from the star \citep{Matzner05, Rafikov09}, 
but if the disk is very massive or, equivalently, has high mass 
accretion rate ${\dot M}\gtrsim 10^{-5} M_\odot$ yr$^{-1}$, then it 
must be optically thick at these separations.
Such high mass accretion rates may be typical during the early, embedded phase of the protostellar evolution, but are unlikely during the classical T Tauri phase. As a result, 
objects formed by disk fragmentation must be born
early and then survive in a disk for several Myrs, which may not be trivial. 
Indeed, gravitational coupling to a massive protoplanetary
disk in which clumps form should naturally drive their rapid radial
migration, potentially resulting in their accretion by the star \citep{Vorobyov06,Machida11, Baruteau11}. Moreover, as condition (1) for disk instability is fulfilled, spiral modes tend to develop in the disk and carry out angular momentum. As a result, the outer parts of the disk will disperse without fragmenting while the inner parts will accrete onto the central star \citep{LaughlinBodenheimer94}. Therefore, in order for a bound object to be able to form by GI and survive, a fragment must both (i) cool (i.e. radiate away the energy induced by the $PdV$ compression work) and (ii) lose angular momentum fast enough to condense out before it is either dispersed out, sheared apart or accreted. Some other processes affecting the survival of collapsing objects have been reviewed in \citep{Kratter11}.


All these mechanisms, radiative cooling, angular momentum redistribution, migration, are by nature strongly non-linear and cannot be captured by linear stability analysis such as the one leading to condition (1), and one must rely on numerical simulations to explore this issue.
Numerical studies of planet or BD formation by gravitational instability, however, have not yet converged to a definite conclusion regarding the viability of this mechanism. Part of the reason for this lies in the quite different numerical schemes employed in these studies. Smoothed Particle Hydrodynamics (SPH) is well suited for modeling self-gravitating systems and following the collapse of forming fragments to very high densities. However, this method has known problems with treating gas dynamics, which plays a key role in disk fragmentation. Grid-based methods have their own problems
and so far many of the numerical studies employing this numerical scheme have reported conflicting results. In contrast, simulations including a detailed thermal evolution of the disk, including stellar irradiation show that these effects severely quench fragmentation \citep{MatznerLevin05}.

A very important source of this ambiguity lies in how the disk cooling is treated in simulations. We saw that the rate at which the disk cools may be a key factor controlling the disk ability to fragment. A number of early studies of disk fragmentation adopted isothermal equation of state for the gas, which is equivalent to assuming that $t_{cool}\to 0$ \citep{Armitage99,Mayer02}. Others assumed less compressible equation of state $P=K\rho^\gamma$ with $\gamma>1$, but this still neglects gas heating in shocks and makes disk easier to fragment \citep{Mayer04}. Not surprisingly, fragmentation has been found to be a natural outcome in such simulations \citep{Boss07}.

To definitively settle the question of disk fragmentation one must be able to follow disk thermodynamics as accurately as possible, and to be able to calculate transport of radiation from the midplane of the disk to its photosphere. Treatment of disk properties at the photosphere is a key issue, since the cooling rate is a very sensitive function of the photospheric temperature $T_{eff}$, see equation (\ref{eq:t_cool}). All existing studies of disk fragmentation incorporating radiation transfer rely on the use of the flux-limited diffusion approximation. This approximation works well in the deep optically thick regions of the disk, as well as in the outer, rarefied optically thin regions. Unfortunately, it is not so accurate precisely at the transition between the two limits, i.e.
at the disk photosphere, where one would like to capture disk properties most accurately.

Treatment of radiation transfer in SPH simulations suffers from similar issues, exacerbated by the fact that in SPH even the very definition of the disk photosphere is very ambiguous and subjective \citep{Mayer07}. There is also a number of other technical issues that might have affected disk fragmentation in some studies, such as the treatment of the equation of state at low temperatures \citep{Boley07,Boss07}, artificial constraints on the disk temperature \citep{Cai10}, and so on. A number of these numerical issues, technical but nevertheless quite crucial, were discussed in detail in \citet{Durisen07} and \cite{Dullemond09}.

Despite these ambiguities, the general consensus that seems to be emerging from numerical and analytical studies of gravitational instability is that disk fragmentation is extremely unlikely in the innermost parts of the disk, where the cooling time is very long compared to the local dynamical time. At the same time, formation of self-gravitating objects may be possible in the outer parts of the disk, beyond 50-100 AU from the star, even though some three-dimensional radiative, gravitational hydrodynamical models claim that even at such distances disk instability is unlikely \citep{Boss06}. In any case, as mentioned above, 
it is still not clear if objects formed by GI can avoid disruption or migration into the star during their residence in a dense protoplanetary disk out of which they formed \citep[e.g.][]{Machida11,Baruteau11,Vorobyov13}.

\section{FORMATION SCENARIOS FOR BROWN DWARFS AND GIANT PLANETS}
\label{sect:formation}

\subsection{Common formation scenario: disk fragmentation}
\label{sect:formation_disk}

Disk fragmentation has been invoked both for the formation of BD's and GP's and thus its pertinence must be examined as a general mechanism for the formation of SSO's. 

As described in the previous section, for a disk to become 
gravitationally unstable and lead eventually to the formation of GP or BD companions, the disk must be massive enough and fulfill the appropriate cooling conditions. 
Some simulations \citep{Stamatellos07, VorobyovBasu06} do predict BD formation at large orbital distances under such conditions, 
provided the disk is massive and extended enough (typically $M_d\gtrsim 0.3\,\,M_\star$, $R_d\gtrsim 100\,(M_\star/\msun)^{1/3}$ AU \citep{Stamatellos09}). 
These simulations, however, lack a fundamental physical ingredient, namely the magnetic field. Indeed, it has been shown by several studies that magnetic fields prevent significant 
mass growth and stabilize the disk, severely hampering fragmentation \citep[e.g.][]{HennebelleTeyssier08,PriceBate09,Commercon10,Machida10}. 
Magnetic fields
drive outflows and produce magnetic breaking, carrying out most of the angular momentum. 
These simulations, however, were conducted with ideal MHD. Subsequent calculations have shown that non-ideal MHD effects \citep[e.g.][]{Dapp12,MachidaMatsumoto11},
misalignment of the field with the rotation axis \citep{HennebelleCiardi09,Joos12,Li13} and the turbulent nature of the velocity field \citep{Joos13,Seifried12,Seifried13}   
 all decrease the efficiency of magnetic braking. However, although large discs can form in these cases, they still remain smaller and less massive than the ones 
produced in pure hydrodynamical simulations. 
Three-dimensional calculations including radiation hydrodynamics, turbulence and complete MHD equations suggest that, for typical observed values of cloud magnetizations and rotation rates, large Keplerian disks do form during the initial core collapse phase
but do not seem to be massive enough to lead to fragmentation {\it (Masson et al., in prep.)}. 
This importance of magnetic field has received strong support from the confrontation of observations carried out at the PdBI with synthetic images derived from simulations of both pure hydro and MDH simulations \citep{Maury10}. It was shown that the disk synthetic images derived from the MHD simulations were similar to the observed ones, with rather compact disks of typical FWHM $\sim 0.2^{''}$-0.6$^{''}$, while the hydrodynamical simulations were producing both too much extended disks and too fragmented (multiple) structures compared with observations. Moreover, as mentioned in \S\ref{sect:obs_disks}, 
observations of isolated disks at the early class 0 stage have revealed compact ($R\sim 20$ AU) disks \citep{Rodriguez05}, 
while so far extended {\it massive} disks prone to fragmentation appear to be rare. Although, as mentioned in \S4.1, better statistics is certainly needed to reach more robust conclusions.
Therefore, although disk fragmentation probably occurs in some disks at the early stages of star formation, 
the fragmentation process is largely overestimated in purely hydrodynamical simulations. Furthermore, as mentioned in \S\ref{sect:GI_physics}, it is not clear
that fragments, if they form in the disk, will be able to cool quickly enough to form bound objects and to survive turbulent motions or rotational shear nor 
that they will not migrate quickly inward and end up been accreted by the star, leading to
episodic accretion events.

Other observational constraints seem to contradict GI in a disk as the {\it main route} for
BD or GP formation. If indeed this mechanism was dominant, most Class 0 objects should have massive disks, a conclusion which does not seem, so far, to be supported by observations. An argument sometimes raised by some proponents of this scenario \citep[e.g.][]{Stamatellos09} is that the lifetime of the massive and extended disks produced in the simulations is too short ($\sim$ a few 10$^3$ yr) for the disks to be observed. This argument, however, does not hold since Class 0 objects last over a significantly longer period, about $\sim 10^5$ yr.
Second of all, these simulations do not include an accreting envelope and thus end up too early compared with realistic situations, yielding an inaccurate diagnostic. 
As mentioned above, more accurate calculations including accreting envelope, magnetic field, turbulence and feedback (e.g. \citet{Seifried13}; Masson et al., in prep.) show the emergence of large centrifugal disks but, for typical observed magnetic flux conditions, the disks do not appear to be massive enough to lead to fragmentation, at least for low mass protostars, the bulk of the
distribution. It is fair to say, however, that at the time these lines are written this important issue is far from being settled and further exploration of this topic is needed. As mentioned above, however,
it is mandatory to include all the proper physics in these studies to reach plausible conclusions. 

Most importantly, the observational constraints arising from the existence of wide BD or GP companions (see \S\ref{BD_cn}+\ref{sec:DI}) strongly argue against GI as an efficient formation mechanism. 
Besides the objects mentioned in \S\ref{BD_cn}, an interesting exemple is the triple system LHS6343 \citep{Johnson11}, with the presence of a BD companion with 
mass $M_C=0.063\, \msun$ to a low-mass star with $M_A=0.37\, \msun$. Assuming disk-to-star standard mass fractions around $\sim 10\%$, the disk around $M_A$ should thus have a mass $M_d\approx 0.04\, \msun$. Although admittedly crude, this estimate shows that there is not enough mass in the disk to have formed the BD companion, even if assuming 100\% disk-to-BD mass conversion efficiency. 
Finally, the statistics arising from direct imaging in \S\ref{sec:DI} revealing, at least so far, the scarcity of wide and massive SSO's in the BD or GP mass range around FGKM-type stars strongly argues against GI as the {\it main} formation
mechanism for BD's or GP's. 
One might invoke the possibility that objects form by GI at large orbits and then migrate inwards closer to the star. This argument, however, does not hold. First, it must be stressed that
the possibility of migration is already included in the general analysis mentioned above. Secondly, if there was a "planetary" population forming by GI and migrating inwards, there should also be a "brown dwarf" population undergoing the same process. But this can be firmly observationally excluded. Finally, if we interpret the existing close-in planet population as objects that
 formed by GI and migrated inwards, this would not explain the aforementioned planet bottom-heavy mass distribution and planet-metallicity correlation.
 Gravitational instability, however, might occur in some particular situations, like for instance in very massive disks around {\it binary systems} \citep{Delorme13}, even though other formation mechanisms are also possible in such situations.
As mentoned in \S\ref{sec:DI}, the combination of ALMA and future direct imaging projects will
definitely help nailing down this issue.

\subsection{Formation scenarios for brown dwarfs}
\label{sect:formation_BD}

In this section, we address formation scenarios which have been specifically suggested for brown dwarfs, besides GI. Note that a more specific examination of the star and BD formation theories can be found in the short review by {\it Hennebelle \& Chabrier} (2011).

\subsubsection{Photoionization}

The fact that the BD mass function is about the same regardless of the presence or not of O stars shows that
halting of accretion by photoionizing radiation \citep{WhitworthZinnecker04} is clearly not a major mechanism for BD formation. Furthermore, as mentioned in \S\ref{IMF}, BD's have been observed in isolated environments, excluding a necessary connection between the presence of photoionizing radiation and BD formation.

\subsubsection{Accretion-ejection }
\label{sec:acc_eje}

In the accretion-ejection scenario \citep{ReipurthClarke01}, BD's are the result of accreting $\sim 1\,\mjup$ stellar embryos, formed by the efficient dynamical fragmentation of a molecular clump,
 which get ejected from the surrounding gas reservoir and thus end up remaining in the substellar domain. The characteristic conditions of this scenario are (1) that dynamical interactions are responsible for the formation of both high-mass (by merging) and low-mass (by ejection) objects, implying that star formation occurs essentially, if not only, in dense cluster environments, (2) that the IMF is determined essentially at the latest stages of the collapse, namely the ultimate gas-to-star conversion,
with no correlation whatsover with the initial core mass function. In fact, prestellar cores do not really exist in this scenario.

The most achieved simulations exploring this scenario are the ones conducted by \cite{Bate12} which,
although still lacking magnetic field, include a treatment of gas heating and cooling. One of the most striking results of these simulations is that the resulting IMF reproduces quite well the Chabrier (2005) IMF (see \S\ref{BD_IMF}). These simulations, however, remain of questionable relevance for most Milky Way (MW) molecular cloud conditions. Indeed, it has been established observationally that star forming molecular clouds in the MW follow the so-called Larson's relations \citep{Larson81}, although with significant scatter, in terms of cloud size vs mean density and velocity dispersion \citep[e.g.][]{HennebelleFalgarone12}:

\begin{eqnarray}
{\bar n} &\simeq& 3\times 10^3\,\left({L\over 1\,{\rm pc}}\right)^{-\eta_d}\,\,\,{\rm cm}^{-3} \\
\sigma_{rms}&\simeq& 0.8\,\left({L\over 1\,{\rm pc}}\right)^{\eta}\,\,\,{\rm km}\,{\rm s}^{-1} 
\end{eqnarray}
with $\eta_d\sim 0.7$-1.0 and $\eta_d\sim 0.4$. Bate's (2012) initial conditions, however, correspond to a 500 $\msun$ cloud at 10 K with size $L=0.4$ pc, thus mean density ${\bar n} \simeq 3.2\times10^4$ cm$^{-3}$ and surface density 
${\Sigma} \approx 10^3$ $\msun$ pc$^{-2}$ (while typical values for MW molecular clouds are around $\sim 60$-100 $\msun$ pc$^{-2}$ \citep{Heyer09}), and Mach number ${\cal M}=14$.
These values correspond to rather extreme cloud conditions, about 4 to 5 times denser and more turbulent
than the aforementioned typical observed ones. Such conditions will strongly favor fragmentation and
dynamical interactions and it is thus not surprising that they produce a significant number of ejected BD embryos. Interestingly enough, these simulations can be directly confronted to observations. Indeed, the simulated cloud mass and size are very similar to the ones of the young cluster NGC1333 \citep{Scholz12b}. The simulations, however, produce a significantly larger number of stars+BD's than the observed ones, with a total stellar mass $\sim 191\,\msun$ against $\sim 50\,\msun$ in NGC1333 \citep{Scholz12b}; interestingly, this echos the aforementioned factors $\sim 4$ in density and velocity. Other inputs in the simulations, for instance the assumption of a uniform initial density profile, the lack of magnetic field, and the underestimated radiative feedback all favor fragmentation and thus probably overestimate the number of protostellar or protoBD embryos
formed in a collapsing core, again favoring dynamical interactions/ejections. Observations in fact tend to suggest that fragmentation within prestellar cores is rather limited, most of the mass of the core ending up in one or just a few smaller cores \citep{Bontemps10, Tachihara02}. 
This again casts doubts on the relevance of such initial conditions to explore star/BD formation under typical Milky Way molecular cloud conditions. At the very least, if indeed BD formation by dynamical ejections might occur under some circumstances, existing simulations severely overestimate the efficiency of this process. In fact, one is entitled to
suspect that the similarity between the IMF produced by the simulations and the one representative of the Galactic field reflects in reality the result of the initial gravoturbulent collapse of the cloud 
rather than the results of dynamical interactions. In which case, the IMF in the simulations should already be largely determined at the begining of the simulation. If this is the case, these
simulations in fact bring support to star/BD formation by gravoturbulent collapse.
At any rate, until simulations with more realistic initial conditions are conducted, the ones performed 
by Bate (2012) cannot be considered as a reliable demonstration of {\it dominant} BD formation by accretion-ejection.

The accretion-ejection scenario faces other important issues. Some of them have already been
mentioned in \S\ref{BD_cn}.
Without being exhaustive, one can add a few more. (1) How can BD's form, according to this scenario, 
in low-density environments? The Taurus cloud for instance,
even though having a stellar density about 3 orders of magnitude smaller than other common clusters, has a comparable abundance of BDs 
\citep{Luhman12}. 
As noted by \cite{Kraus11}, the low stellar density ($\lesssim$ 5 stars /pc$^2$) and low 
velocity dispersion of Taurus members 
($\sigma \sim 0.2$ km s$^{-1}$), indicate that there is no small {\it N}-body clusters from which stars or BDs could have been ejected, as advocated in the accretion-ejection scenario.
The fact that BD's form as efficiently in such low-density environments strongly argues against dynamical interactions.
(2) Various observations show average dispersion velocities of prestellar cores of about $\langle \sigma \rangle \sim 0.4$ km s$^{-1}$ \citep{Andre09, Walsh07, Muench07, Gutermuth09, Bressert10}. This indicates a typical collision timescale between prestellar cores significantly larger than their dynamical timescale, suggesting little dynamical evolution during prestellar core formation. (3) How can this scenario explain 
 the observed similarity between the CMF and the IMF, if indeed such a similarity is confirmed? 

All these observational constraints (notably the observed small velocity dispersion between protostar/BD's) severely argue against the accretion-ejection scenario for BD or star formation as a {\it dominant} mechanism, except possibly for the most massive stars, which represent only a very small fraction of the stellar population.

\subsubsection{Gravoturbulent fragmentation }
\label{sec:gravo}

In this scenario, large-scale turbulence injected at the cloud scale by various sources 
cascades to smaller scales by shocks and generates a field of density fluctuations down to the dissipative scale \citep[e.g.][]{MacLowKlessen2004}. Overdense regions inside which gravity overcomes all other sources of support collapse and form self-gravitating cores which isolate themselves from the surrounding medium. In this scenario, the CMF/IMF is set up by the spectrum of turbulence, at the very early stages of the star formation process. 
The first theory combining turbulence and gravity was proposed by \cite{PadoanNordlund02} but has been shown to suffer from various caveats (e.g. \cite{McKeeOstriker07, HC11}). 
A different theory was derived more recently by \cite{HC08,HC09, HC13} and \cite{Hopkins12} (see chapter by {\it Offner et al.}). These latter theories nicely reproduce the C05 IMF down to the least massive BDs, for appropriate (Larson like) conditions. A common and important feature
 of these theories is that it is inappropriate to use the average thermal Jeans mass as an estimate of the characteristic mass for fragmentation. This would of course preclude significant BD formation. Indeed, in both Hennebelle-Chabrier and Hopkins theories, the spectrum of collapsing prestellar cores strongly depends on the Mach number, shifting the low-mass tail of the IMF to much smaller scales than naively expected from a purely gravitational Jeans argument. The theories indeed yield a reasonably accurate number of pre-BD cores for adequate molecular cloud like conditions and Mach values (${\cal M}\sim 3$-8). Densities required for the 
 collapse of BD-mass cores, $\sim 10^7$-$10^8$ g cm$^{-3}$ (see \S\ref{BDcore}), are indeed produced by turbulence induced shock compression (remembering that shock conditions imply density enhancements $\propto{\cal M}^2$). 
Note that a similar scenario, taking into account the filamentary nature of the star forming dense regions, was developed by \cite{Inutsuka2001} (see chapter by {\it Andr\'e et al.}). 

 Support for the gravoturbulent scenario arises from several observational facts. (1) It explains, within the framework of the same theory, the observed mass spectra of both {\it unbound} CO-clumps and {\it bound} prerestellar cores \citep{HC08}; (2) it implies that the IMF is already imprinted 
 in the cloud conditions (mean density, temperature and Mach number), naturally explaining the resemblance of the CMF with the IMF; (3) it relies on one single "universal" parameter, namely the velocity power spectrum index of turbulence, which explains as well the Larson's relations for molecular clouds.

A major problem against the gravoturbulent scenario to explain the {\it final} IMF would arise if the cores were fragmentating significantly into smaller pieces during their collapse. As mentioned in the previous section, however, observations tend to show that
such fragmentation is rather limited. Numerical simulations of collapsing dense cores indeed show that radiative feedback and magnetic fields drastically reduce the fragmentation process \citep[e.g.][] {Krumholz07, Offner2009, Commercon10, Commercon11, Hennebelle11,Seifried13}. 
The analysis of simulations aimed at exploring the CMF-to-IMF conversion \citep{Smith09} also show a clear correlation between the initial core masses and the final sink masses up to a few local freefall times \citep{CH10}, suggesting that, at least for the
bulk of the stellar mass spectrum, the initial prestellar cores do not fragment into many objects, as indeed suggested by the CMF/IMF similarity. But the most conclusive support for BD formation by gravoturbulent fragmentation comes from the emerging observations of isolated proto-BD's and of
the pre-BD core Oph B-11, with densities in agreement with the afortementioned value (see \S\ref{BDcore}).

\subsubsection{Formation of binaries}
\label{bin_fn}

Newly formed stars must disperse a tremendous amount of angular momentum in condensing through more than 6 orders of magnitude in radius \citep{Bodenheimer95}. Besides magnetic braking, binary formation offers a  convenient way
to redistribute at least part of this excess angular momentum. 

As mentioned in \S\ref{BD_cn}, several binaries have been observed at large {\it projected} separations ($>500$ AU) with masses down to $\sim 5\,\mjup$. \cite{Looney00} showed that multiple systems in the Class 0 and I phases are prevalent on large spatial scales ($\gtrsim$ 1000 AU) 
while binary systems at smaller scales ($\lesssim$ 500 AU) seem to be quite sparse \citep{Maury10}.
Whether this lack of small scale binaries at the Class 0/I stages is real or stems from a lack of observations with sufficient resolution or sensitivity, however, is still unclear so far.
In any case, prestellar and protoplanetary disks do not extend out to thousand AU's so
formation of such wide BD binaries by disk GI seems to be clearly excluded. In contrast, such distances can be compared with the characteristic sizes of starless prestellar core envelopes, $\sim 10^4$ AU \citep{Menshchikov10}. The existence of these wide systems thus suggests
that prestellar core fragmentation into binaries might occur at the very early stages (typically at the class 0 stage or before) of the collapse. Indeed, there are some observational evidences for binary fragmentation at this stage \citep{Looney00,Duchene03}. Moreover, the similarity, at least in a statistical sense, between some of the star and BD multiplicity properties (see \S\ref{BD_cn}) suggests that most BD companions form similarly as stellar companions. This hypothesis has received theoretical 
and numerical support
form the recent work of \cite{JumperFisher13}. These authors show that BD properties, including the BD desert, wide BD binaries and the tendency for low-mass (in particular BD) binaries to be more tightly bound than stellar binaries, arguments often used as an evidence for distinct formation mechanisms for BD's, can de adequately reproduced by angular momentum scaling during the collapse of turbulent clouds. 
Indeed, the early stages of cloud collapse/fragmentation and core formation are characterized by the formation of puffy disc-like structures which keep accreting material 
from the surrounding core envelope. These structures, however, are not relaxed and differ from structures purely supported by rotation, characteristic of relaxed, equilibrium discs.
 Such "pseudo-discs" may fragment (as the result of global non-linear gravitational instability) during the (first or second) collapse of the prestellar core and end up forming (wide or tight) binaries, 
possibly of BD masses \citep{BonnellBate94,Machida08}. 
The occurence of such fragmentation at the cloud collapse stage has been found in radiation-hydrodynamics simulations \citep{Commercon10, Peters11, Bate12}. 
They remain to be explored in the context of resistive MHD before more definitive conclusions can be reached concerning this important issue. 
As just mentioned, such fragmentation into binaries occurs at the {\it very early stages of the core collapse}, during the main accretion phase. 
Whether it occurs dominantly by redistribution of angular momentum during the collapse itself or by global instability in the mass loaded growing pseudo-disk remains so far
 an open question and a clear-cut 
distinction between the two processes is rather blurry at this stage. 
It must be kept in mind, however, that the conditions for the formation {\it and} survival of bound fragments in a disk are subject to very restrictive constraints, as examined in \S4.2.

\subsection{Formation scenarios for giant planets: core accretion}
\label{sect:formation_GP}

\subsubsection{Core accretion mechanism}
\label{sect:formation_CA}

In the core accretion scenario gas-giant planets form in two steps. 
First, a solid core is accumulated by accretion of planetesimals.  The growing
core attracts a hydrostatic envelope of protoplanetary disk gas, extending from
the surface out to the Bondi radius where the sound speed equals the escape
speed of the core. Since accreting planetesimals not only increase the core
mass but also release gravitational energy, the energy released at the core
surface must be transported through the envelope, making the latter
non-isothermal. The mass of the envelope increases with the core mass faster
than linearly, so that at some point the envelope starts to dominate the gravitational
potential. Hydrostatic solutions cease to exist beyond a critical core mass
which depends mainly on the accretion rate of solids and on the gas opacity. As
a result, core starts accreting mass on its own thermal timescale, turning
itself into a giant planet.

\citet{Perri74} assumed the hydrostatic envelope to be fully convective and found critical core masses in
excess of 100 Earth masses; they used this result to support the alternative gravitational instability scenario for giant planet
formation.  \citet{Mizuno80} allowed both for convective and radiative regions
in the envelope and constructed the equilibrium model using gas opacities
depending on the density and temperature, with a stepwise constant dust opacity
below the dust sublimation temperature. The radiative solution drastically
reduces the critical core mass to approximately 10 $\mearth$, in agreement with
constraints from the gravitational potentials of Jupiter and Saturn
\citep{Guillot05}, for an assumed mass accretion rate of $10^{-6}\, \mearth$
per year.  Planetary envelopes can still be fully convective at very high
planetesimal accretion rates \citep{Wuchterl93,Ikoma01,Rafikov06},  typical for
cores at $\sim$AU from the star, but this is unlikely to be an issue at large
separations \citep{Rafikov06}.

A generic property of the hydrostatic envelope is that its mass is inversely
proportional to the luminosity (and hence to the mass accretion rate) and to
the opacity \citep{Stevenson82}. The resulting critical core mass roughly
scales as $M_c\propto {\dot M}^{1/4}$ \citep{Ikoma00}. Hence the growth rate of
the core partially determines the critical core mass. Beyond the critical core
mass the envelope will emit more energy than provided by the gravitational
potential release of the accreted solids and undergo run-away contraction on
the Kelvin-Helmholtz time-scale \citep{BodenheimerP86}.  
Interestingly, the concept of a critical core mass seems to be supported observationally by the mass-radius relationship of the recently detected Kepler low-mass transit planets \citep{Lissauer11a,Lissauer11b,Carter2012,Ofir2013}, bearing in mind the uncertainties in Kepler objects mass determinations. 
Indeed, while planets above about $\sim$$6$ $\mearth$ seem to have a radius requiring a substantial ($\gtrsim$$10\%$ by mass) gaseous H/He envelope, {\it so far} planets below this mass have a radius consistent with a much lower, or even negligible gas {\it mass} fraction.
Even the rather extended gaseous envelope of the lowest transiting planet discovered so far, KOI-314c ($1.0_{-0.3}^{+0.4}M_{\oplus}$) \citep{Kipping14}, with a radius $R_{env}\sim 7_{-13}^{+12}\%$ of the planet's radius,
represents a negligible fraction ($\lesssim 3\%$) of its mass.

\subsubsection{Time-scale for accumulation of the core}

In the classical core accretion scenario, the core grows by accreting
planetesimals. The Hill radius
\begin{equation}
  R_{\rm H} = [G\,M_p/(3\,\Omega^2)]^{1/3} = \frac{1}{p} R_{\rm p}
\end{equation}
denotes the maximal distance over which the core can deflect planetesimals
which pass by with the (linearized) Keplerian shear flow. Here
$p=(9M_\star/4\pi\rho a^3)^{1/3}\ll 1$ is a parameter that depends on 
semi-major axis
$a$ and material density $\rho$ for a given stellar mass $M_\star$
\citep{Goldreich+etal2004}. An additional random component to the particle
motion will reduce the scattering cross section of the core to below $\sim
R_H^2$.

The gravitational radius of the core denotes the maximum impact 
parameter at which a planetesimal approaching with velocity $v$ gets accreted 
by a core with radius $R$ at closest approach. This distance is given by
\begin{equation}
  R_{\rm G} = R \sqrt{ 1 + \frac{v_{\rm e}^2}{v^2} } \, ,
\end{equation}
where $R$ is the radius of the core, $v_{\rm e}$ the escape speed at
the surface and $v$ the relative approach speed. In dynamically
cold discs planetesimals enter the Hill sphere of the core with the 
Hill speed $v_{\rm H}=\Omega R_{\rm H}$. In that case
the gravitational radius can be simplified as
\begin{equation}
R_{\rm G} \approx \sqrt{p}\, R_{\rm H}.
\end{equation}
A small fraction of the planetesimals which enter the Hill sphere of the core actually collide with it. The
mass accretion rate is ultimately set by the gravitational radius and the scale
height of the planetesimals. The random motion of planetesimals enter both
these quantities and should ideally be calculated self-consistently using an
{\it N}-body approach \citep[e.g.][]{Levison+etal2010}.

Assuming that the random particle motion is similar to $\Omega R_H$ --- 
the Keplerian velocity shear across the Hill radius, one can find that the core grows at the rate \citep{Dones93}
\begin{equation}
dR/dt\sim p^{-1}\Sigma_s\Omega/\rho\approx 50\,\,{\rm m\, yr}^{-1}(r/{\rm AU})^{-2},
\label{growth1}
\end{equation}
assuming surface density of solids typical for the MMSN
($\sim 30$ g cm$^{-2}$ at 1 AU).
This corresponds to the mass growth rate of order 
$10^{-6}\,\mearth/$year at 5 AU where Jupiter presumably formed
\citep{Dodson09}. This leads to core formation in $10^7$ years, in
marginal agreement with the observed life times of protoplanetary discs
\citep{Haisch+etal2001}, but the time-scale can be brought down by assuming a
planetesimal column density that is 6-10 times the value in the MMSN
\citep{Pollack+etal1996}.
Note that eqn. (11) assumes uniform planetesimal surface density 
near the embryo, an assumption which may be violated if the embryo is
capable of clearing a gap in the planetesimal population around its 
orbit \citep{TanakaIda97, Rafikov01, Rafikov03a}, which would slow down 
its growth \citep{Rafikov03b}.

While the agreement between the core mass (or more exactly the total amount of heavy material) of Jupiter and the mass accretion
rate necessary to form the core is a success for the core accretion model, the
mass accretion rate scales roughly as the semi major axis to the inverse  second power in accordance with equation (11). Hence core formation beyond 5 AU is impossible in the million-year
time-scale of the gaseous protoplanetary disc. This conflicts both with the
high gas contents of Saturn, the small H/He envelope of Uranus and Neptune and
the observations of gas giant exoplanets in wide orbits beyond 10 AU
\citep{Marois+etal2008}.

This discussion makes it clear that gas giants can form by CA at large
separations only if the core formation timescale can be reduced dramatically.
Ironically, acceleration of the core growth makes it harder to reach the
threshold for CA, because the critical core mass is inversely proportional to
the core luminosity, which is predominantly derived from the heat released in
planetesimal accretion. Nevertheless, one can show \citep{Rafikov11} that
higher planetesimal accretion rate $\dot M$ still reduces the total time until
CA sets in, even though the critical core mass is larger and is more difficult
to achieve.

\subsubsection{Fragment accretion}
\label{frag}

One pathway to rapid core formation is to accrete planetesimal fragments damped
by the gas to a scale height lower than the gravitational accretion radius
\citep{Rafikov04}. 

Numerical and analytical studies of planetesimal dynamics generically find that
soon after a dominant core emerges it takes over dynamical stirring of the
surrounding planetesimals, increasing their random velocities to high values.
Physical collisions between planetesimals then lead to their destruction rather
than accretion, giving rise to a fragmentation cascade that extends to small
sizes. \citet{Rafikov04} demonstrated that random velocities of particles in a
size range $1-10$ m are effectively damped by gas drag against excitation by
the core gravity. Gas drag also 
does not allow such small fragments to be captured in mean motion 
resonances with the embryo  \citep[e.g.][]{Levison+etal2010}. This velocity damping is especially effective in the vertical
direction, allowing small debris to settle into a geometrically thin disk with
thickness below the maximum impact parameter, still resulting in planetesimal
accretion  ($\sim$$p^{1/2} R_{\rm H}$). From the accretion point of
view the disk of such planetesimal debris is essentially two-dimensional,
resulting in the highest mass accretion rate possible in the absence of any
external agents such as described next (see \S\ref{Pebble} below):  

\begin{equation}
dR/dt\sim p^{-3/2}\Sigma_s\Omega/\rho\approx 1\,\,{\rm km\, yr}^{-1}(r/{\rm AU})^{-3/2}.
\end{equation}
This formula assumes that most of the surface density of solids 
$\Sigma_s$ ends up in small ($\sim 10$ m) fragments dynamically 
cooled by gas drag.
Because of the dramatically reduced vertical scale height of the 
fragment population, this rate is higher than the previously quoted (\ref{growth1}) for the velocity dispersion $\sim \Omega R_H$ and decreases somewhat more slowly with the distance (as $r^{-3/2}$). As a result, a Neptune size core can be formed at $30$ AU within several Myrs.

A key ingredient of this scenario is the collisional grinding of
planetesimals, which allows them to efficiently couple to gas and become
dynamically cold, accelerating core growth. Fragmentation plays a role of an
agent that transfers mass from a dynamically hot mass reservoir (population of
large planetesimals, accreted at low rate because of relatively weak focussing)
to dynamically cold debris. Even though at a given moment of time the latter
population may have lower surface density than the former, it is accreted at
much higher efficiency and easily dominates core growth.
The importance of fragment accretion for accelerating core growth has been confirmed in recent coagulation simulations by \cite{KenyonBromley09} (see also \cite{Levison+etal2010}).

In the framework of this scenario \citet{Rafikov11} sets a constraint on the
distance from the star at which gas giants can still form by CA in a given
amount of time (several Myrs). By adopting the maximally efficient accretion of
small debris confined to a two-dimensional disk he showed that giant planet
formation can be extended out to 40-50 AU in the MMSN and possibly even further
in a more massive planetesimal disk (out to 200 AU in the marginally
gravitationally unstable gaseous protoplanetary disk). This estimate makes some
rather optimistic assumptions such as that most of planetesimal mass is in
small (1-100 m) objects (e.g. as a result of efficient fragmentation), and that
seed protoplanetary embryos with sizes of at least several hundred km can be
somehow produced far from the star within several Myrs. Thus, it becomes very difficult to
explain origin of planets beyond 100 AU by CA. This issue is examined below.

\subsubsection{Pebble accretion}
\label{Pebble}

The arguably most fundamental particles to accrete are those forming directly
by coagulation and condensation as part of the planetesimal formation process.
Several particle growth mechanisms predict that particles will stop growing
efficiently when they reach pebbles of millimeter and centimeter sizes.

While small dust grains grow as fluffy porous aggregates, they are eventually
compactified by collisions around mm sizes and enter a regime of bouncing
rather than sticking \citep{Zsom+etal2010}. This bouncing barrier maintains a
component of small pebbles which can feed a growing core. In the alternative
mechanism where particles grow mainly by ice condensation near evaporation
fronts, particles sediment out of the gas when reaching pebble sizes. This
reduces their growth rate drastically as they must collectively compete for the
water vapour in the thin mid-plane layer \citep{RosJohansen2013}. Icy particles
may be able to retain their fluffy structure against compactification and hence
avoid the bouncing barrier \citep{Wada+etal2009}, but even under perfect
sticking particles grow more slowly with increasing mass and spend significant
time as fluffy snow balls whose dynamics is similar to compact pebbles
\citep{Okuzumi+etal2012}.

The dynamics of pebbles in the vicinity of a growing core is fundamentally
different from the dynamics of a planetesimal or a planetesimal fragment.
Pebbles are influenced by drag forces which act on a time-scale comparable to or
shorter than the orbital period, which is also the characteristic time-scale
for the gravitational deflection of particles passing the planet with the
Keplerian shear. The gravity of the core pulls the pebbles out of their
Keplerian orbits. The resulting motion across gas streamlines leads to
frictional dissipation of the kinetic energy of the particles and subsequent
accretion by the core. While planetesimals are only accreted from within the
gravitational radius, which is a small fraction of the Hill radius, pebbles are
in fact accreted from the {\it entire} Hill sphere.

The potential of drag-assisted accretion was explored by
\cite{WeidenschillingDavis1985} who showed that particles below 1-10 meters in
size experience strong enough friction to avoid getting trapped in mean motion
resonances with the growing core. \cite{Kary+etal1993} confirmed these results
but warned that only a small fraction (10-40\%) of the drifting pebbles are
accreted by the core due to their radial drift. This concern is based on the
assumption that pebbles form once and are subsequently lost from the system by
radial drift. This loss of solids would be in disagreement with observations of
dust in discs of Myr age \citep{Brauer+etal2007}. The solution to the drift
problem may be that pebbles are continuously formed and destroyed by
coagulation/fragmentation and condensation/sublimation processes
\citep{Birnstiel+etal2010,RosJohansen2013}.  Hence, any material that is not
immediately accreted by the one or more growing cores will get recycled into
new pebbles that can then be accreted.

\cite{JohansenLacerda2010} simulated the accretion of small pebbles onto large
planetesimals or protoplanets. They found that pebbles with friction times less
than the inverse orbital frequency are accreted from the entire Hill sphere, in
agreement with the picture of drag-assisted accretion. \cite{OrmelKlahr2010}
interpreted the rapid accretion as a sedimentation of particles towards the
core. \cite{JohansenLacerda2010} also observed that the pebbles arrive at the
protoplanet surface with positive angular momentum, measured relative to the
angular momentum of the disc, which leads to prograde rotation in agreement
with the dominant rotation direction of the largest asteroids.

The dependence of the pebble accretion rate on the size of the core was
parameterized in \cite{LambrechtsJohansen2012}. Below a transition core mass of
around 0.01 Earth masses pebbles are blown past the core with the sub-Keplerian
gas. Pebbles are captured from within the Bondi radius of the sub-Keplerian
flow, which is much smaller than the Hill radius. With increasing core mass the
Bondi radius grows larger than the Hill radius, starting the regime of
efficient pebble accretion. Growth rates in this regime are 1,000 times higher
than for accretion of large planetesimals with velocity dispersion of order $\Omega R_H$ (see eqn. (11)), at 5 AU, and 10,000 faster, at 50
AU, depending on the degree of sedimentation of pebbles. Thus it is possible to
make cores of 10 $\mearth$ even in the outer parts of protoplanetary discs,
which may explain observed gas giant exoplanets in wide orbits
\citep{Marois+etal2008}.

The accretion rate of pebbles depends on the (unknown) fraction of the solid
mass in the disk which has grown to pebble sizes. There is observational
evidence for large pebble masses in some protoplanetary discs -- the disc
around TW Hya should for example contain 0.001 $M_\odot$ of approximately
cm-sized pebbles to match the emission at cm wavelengths \citep[]{Wilner+etal2005}.
Such high pebble masses are surprising, given the efficiency for turning
pebbles into planetesimals via various mechanisms for particle concentration
(discussed in the next session). However, if planetesimals form through
streaming instabilities, then the simulations show that only around 50\% of the
available pebble mass is incorporated into planetesimals \citep[]{johansen2012}. The remaining pebbles are unable to concentrate in the gas flow, due to
their low mass loading in the gas, and are available for a subsequent stage of
pebble accretion onto the largest planetesimals.

The range of optimal particle sizes for pebble accretion falls significantly
below the planetesimal fragments described above. Direct accretion is most
efficient for particles which couple to the gas on time-scales from the inverse
Keplerian frequency down to 1\% of that value. That gives typical particle
sizes in the range of cm-m in the formation zone of Jupiter and mm-cm for
planet formation in wide orbits, due to the lower gas density in the outer
parts of protoplanetary discs. However, \cite{MorbidelliNesvorny2012} showed
that even boulders of 10 meters in size can be accreted with high efficiency,
following a complex interaction with the core.

Although pebbles are not trapped in mean motion resonances, their motion is
sensitive to the pressure profile of the gas component. The global pressure
gradient drives radial drift of pebbles towards the star, but any local
pressure maximum will trap pebbles and prevent their drift to the core.
\cite{PaardekooperMellema2006} showed that planets above approximately 15
$\mearth$ make a gap in the pebble distribution and shut off pebble accretion.
The shut off is explained in the analytical framework of
\cite{MutoInutsuka2009} as a competition between radial drift of particles and
the formation of a local gas pressure maximum at the outside of the planetary
orbit. The core mass is already comparable to the inferred core masses of
Jupiter and Saturn when the accretion of pebbles shuts down.
\cite{MorbidelliNesvorny2012} suggested that the termination of pebble
accretion will in fact be the trigger of gravitational collapse of gas.

\subsubsection{Implications of rapid core accretion models}

In the pebble accretion scenario, cores of 10 $\mearth$ or more can form at any
location in the protoplanetary disc, provided that a sufficient amount of material
is available. This can now be confronted with observational constraints from
exoplanet surveys. Direct imaging surveys now have good upper limits for the
occurrence rate of Jupiter-mass planet in wide orbits (\S\ref{sec:DI}).
While many aspects of the observations and the luminosity of young planetary
objects are still unknown, it seems that gas-giant planets are rare beyond 25
AU (with $<23\%$ of FGKM stars hosting gas giants in such orbits). This can be
interpreted in three ways: (1) most of solids (pebbles, planetesimals) in
protoplanetary discs reside in orbits within 20 AU, (2)
wide orbits are mostly populated by ice giants as in the solar system, and (3)
gas-giant planets do form often in wide orbits but migrate quickly to the inner
planetary system or are parts of systems which are unstable to planet-planet
interactions shortly after their formation.
Regarding point (1), observations of protoplanetary discs in mm and cm
wavelengths often find opacities that are consistent with large populations of
pebbles \citep[e.g.][]{Wilner+etal2005,
Rodmann+etal2006,Lommen+etal2007}.
The fact that these pebbles are only partially aerodynamically coupled to the
gas flow, an important requirement for pebble accretion, was confirmed in the
observations of a pebble-filled vortex structure in the transitional disc
orbiting the star Oph IRS 48 \citep{vanderMarel2013}.
 However, \cite{Perez+etal2012} showed that particles are generally larger closer to the star, in
agreement with a picture where the outer regions of protoplanetary discs are
drained of pebbles by radial drift. 
Regarding point (2), namely that wide orbits may be dominated by ice
giants, it is an intriguing possibility, but its validation requires a
better understanding of why ice giants in our own solar system did not
accrete massive gaseous envelopes.
Finally regarding point
(3), it is important to keep in mind that massive planets in wide orbits have a
large gravitational influence which can lead to a disruption of the system. The
HR8799 system may owe its stability to a 4:2:1 resonance \citep{FabryckyMurrayClay2010}, which indicates that a significant fraction of similar
systems, less fortunately protected by resonances, may have been disrupted
early on.

\subsubsection{Formation of the seeds}

Efficient accretion of pebbles from the entire Hill sphere requires cores that
are already more massive than 0.1-1\% of an Earth mass. 
Massive seed embryos are needed also in the rapid fragment accretion scenario  \citep{Rafikov11} (see \S\ref{frag}).
Below this transition
mass pebbles are swept past the core with the sub-Keplerian wind. One can
envision three ways to make such large seeds: (1) direct formation of very
large planetesimals, (2) run-away accretion of planetesimals, (3) inefficient
pebble accretion from large Ceres-size, the largest asteroid in the asteroid belt, planetesimals to the transition mass.

The growth from dust to planetesimals is reviewed in two other chapters in this
book ({\it Testi et al.}, {\it Johansen et al.}). Dust and ice grains grow
initially by colliding and sticking to particle sizes which react to the
surrounding gas flow on an orbital time-scale. These pebbles and rocks, ranging
from mm to m in size, experience strong concentration in the turbulent gas
flow. Particles get trapped in vortices
\citep{BargeSommeria1995,KlahrBodenheimer2003} and in large-scale pressure
bumps which arise spontaneously in the turbulent gas flow
\citep{Johansen+etal2009a}. Particles can also undergo concentration through
streaming instabilities which occur in mid-plane layers of unity dust-to-gas
ratio, as overdense filaments of particles catch more and more particles
drifting in from further out
\citep{YoudinGoodman2005,JohansenYoudin2007,Johansen+etal2009b,BaiStone2010}.
\cite{YoudinGoodman2005} performed a linear stability analysis of the
equilibrium flow of gas and particles in the presence of a radial pressure
gradient. The free energy in the streaming motion of the two components -
particles orbiting faster than the gas and drifting inwards, pushing the gas
outwards - forms the base of an unstable mode with growth rate depending on the
particle friction time and on the local dust-to-gas ratio. Higher friction time
and higher dust-to-gas ratio lead to shorter growth time. The growth time
decreases by 1-2 orders of magnitude around a dust-to-gas ratio of unity, which
shows the importance of reaching this threshold mass-loading by sedimentation.

The streaming instability can concentrate particles up to several thousand
times the local gas density in filamentary structures \citep{Johansen+etal2012}.
When particle densities reach the Roche density, the overdense filaments
fragment gravitationally into a number of bound clumps which contract to form
planetesimals \citep{Johansen+etal2007,Johansen+etal2009b,Nesvorny+etal2010}.
The characteristic sizes of planetesimals formed by a gravitational instability
of overdense regions seems to be similar to or smaller than the dwarf planet
Ceres, the largest asteroid in the asteroid belt. Smaller planetesimals form as
well, particularly in high-resolution simulations \citep{Johansen+etal2012}.
However, for pebble accretion the largest planetesimal is the interesting one,
since the accretion efficiency increases strongly with increasing size.
Nevertheless, a planetesimal with the mass of Ceres is still a factor 10 to 100
too light to undergo pebble accretion from the full Hill sphere. Hence
suggestion (1) above, the direct formation of very massive planetesimals, is
not supported by the simulations. A more likely scenario is a combination of
(2) and (3), i.e. that a large Ceres-mass planetesimal grows by a factor 100 in
mass by accreting other planetesimals through its gravitational cross section
and by accreting pebbles from the Bondi radius. This intermediate step between
planetesimals and the seeds of the cores could act as a bottleneck mechanism
which prevents too many planetesimals from reaching the pebble accretion
stage.  Hence only a small number of cores are formed from the largest and
luckiest of the planetesimal population, in agreement with the low number of
giant planets in the solar system.
The growth from 100-km-scale planetesimals to 1000-km-scale planetary seeds,
capable of rapid pebble accretion from the entire Hill sphere, has nevertheless
not been modelled in detail yet and represents an important priority for the future.
The formation of a too high number of competing cores could drastically reduce the
efficiency of this growth path. \cite{NesvornyMorbidelli2012} nevertheless
argued for an initial number of no more than 10 growing cores which were
subsequently reduced in number by collisions and ejections.





\section{Brown dwarf vs giant planets. Questioning the IAU definition}
\label{def}

\subsection{\textbf{Hot start vs cold start}} 
\label{hot_cold}

\cite{Marley07} suggested that young giant planets formed by core accretion in a protoplanetary disk should
have a much lower entropy content at young ages than objects of same mass and age formed by gravitational collapse, either from a prestellar core or by disk instability. Consequently,
the first type of objects should be significantly smaller, cooler and (about 100 times) fainter at young ages than
the second ones. This gave rise to the so-called "cold start"
   observational signature, characteristic of  young GP's formed by core accretion, vs the "hot start" one, typical of young objects, GP's or BD's, formed by collapse to distinguish these objects from their distinct formation mechanisms. This suggestion, however,
  directly relies on two assumptions. First of all, the assumption that GP's formed by core accretion
  have a low entropy content implies that all the energy of the accretion shock through which most of the planetary mass is processed is radiated away, leaving the internal energy content of the nascent planet unaffected. This is characteristic of a so-called supercritical shock. 
In that case the radiative losses at the accretion shock act as a sink of entropy.
  This assumption, however, has never been verified. In fact, a proper treatment of the accretion shock at the onset of a prestellar core formation, the so-called second Larson's core, shows that the
  shock in that case is {\it subcritical}, with essentially all the energy from the infalling material been absorbed by the stellar embryo \citep{Vaytet13,Tomida13,Bate14}. Although the two formation conditions differ in several ways, they share enough common processes to at least question the assumption of a supercritical shock for planet formation. 
  The second underlying
  assumption about this scenario is the assumption that BD's form with a high entropy content. Again, such an assumption is not necessarily correct. Initial conditions for brown dwarf formation are rather
  uncertain. Figure 2 (see also \cite{Mordasini12a, Mordasini13, SpiegelBurrows12}) compares the early evolution of the luminosity for a 5 M$_{Jup}$ object under several assumptions. The blue solid and dot-dash lines portray giant planet early evolution tracks kindly provided by C. Mordasini assuming that either the accretion energy is entirely converted to radiation, i.e. a supercritical shock condition as in \cite{Marley07}, or that this energy is absorbed by the planet, i.e. a subcritical shock with no radiative loss at the shock.
  The former case yields a low entropy content thus a low luminosity at the end of the accretion shock while in the second case the luminosity slowly decreases from its value at the end of the shock, two orders of magnitude brighter, for several Myr's. The red long-dash and short-dash lines correspond to the early evolution of a brown dwarf \citep{Baraffe03} assuming an initial radius $R_i\sim 8\,R_{Jup}(\simeq 0.8\,R_\odot$) (short-dash) or $R_i\sim 1.6\,R_{Jup}$ (long dash). 
  This corresponds to specific entropies ${\tilde S}\simeq 1.1\times 10^9$ and $\simeq 8.0\times 10^8$ erg g$^{-1}$ K$^{-1}$, respectively. 

\begin{figure}[!ht] 
\epsscale{1.0}
\plotone{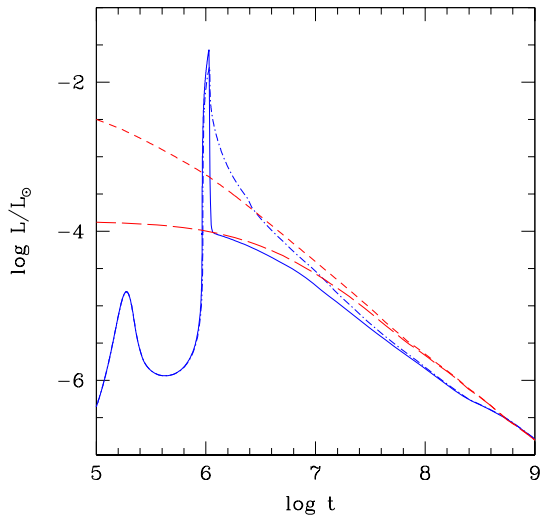}
\caption{\small Early evolution of a 5 Jupiter-mass object according to different formation scenarios.
Blue: object formed by core accretion assuming either a supercritical shock (solid line) or a subcritical shock (dash-dot line) at the end of the accretion process, after \cite{Mordasini12a}.
Red: object formed by gravitational collapse for two arbitrary initial conditions (see text), after \cite{Baraffe03}
}  
\label{fig_hot}
\end{figure}

  As seen in the figure, depending on the outcome of the accretion shock episode in one case and on the initial radius (thus entropy content) in the second case, and given the long Kelvin-Helmholtz timescale for such low-mass objects (several Myrs) young planets formed by core accretion can be as bright 
  and even brighter than young BD's for several Myrs. 
  (Magneto)Radiation-hydrodynamics calculations of the collapse of prestellar cores \citep{Tomida13, Vaytet13} seem to exclude the above rather extreme "cold start" initial condition for a proto-BD, even
   though calculations have not been performed yet for such low masses. They suggest initial entropy contents and radii closer to the hot-start case, the outer region of the protostellar core been
heated up by the shock and attaining a higher entropy. 
Hot-start conditions for core-accretion GP's, probably in between the two above extreme cases, 
however, are presently not excluded. Interestingly enough, measured temperatures and luminosities of some directly-imaged exoplanets, notably $\beta$ Pic b and  $\kappa$ And b, are consistent with hot-start like
conditions and seem, so far, to exclude the coldest range of initial conditions for these objects \citep{Currie13,marleau2013}.
  Therefore, at least in the absence of a better knowledge of the accretion shock condition at the end of the core accretion process, the cold start - hot start argument does not provide a reliable diagnostic to distinguish CA from GI formed objects. For sure there is no one-to-one correspondence 
  between cold start vs. hot start conditions and core accretion vs. disk or core collapse.

\subsection{\textbf{Deuterium burning}} 

The distinction between brown dwarfs and giant planets has become these days a topic of intense debate. In 2003,
the IAU has adopted the deuterium-burning (DB) minimum mass, $\sim 10\,\mjup$, as the official distinction between the two types of objects.
We have discussed this limit in previous reviews \citep{Chabrier03, Chabrier07} and shown that it does not rely on any robust physical justification and is a pure semantic definition.
Deuterium burning has no impact on star formation and a negligible impact on stellar/BD evolution \citep{CB00}. This is in stark constrat with the lifetime impact of hydrogen-burning, making
H-burning a genuine physical mechanism distinguishing objects in nuclear equilibrium for most of their lifetime, defined as stars, from objects which lack significant support against gravitational contraction and keep contracting for ever since their birth, defined as brown dwarfs. 

As mentioned in \S\ref{sec:DI}, 
one of the strongest arguments against DB to distinguish planets from brown dwarfs is 2M~1207~b \citep{chauvin2005}, which is a $\sim$4~$M_{\rm Jup}$ companion to a $\sim$20~$M_{\rm Jup}$ brown dwarf. Thus, the companion is firmly in the non-DB mass regime, but with a system mass ratio of 20\%, the couple appears to be best described as an extension of the brown dwarf binary population rather than a planetary system. In contrast, it is presently not excluded that genuine planets formed by core-accretion, characterized by a significant heavy element enrichment, reach
masses above the DB limit and thus ignite D-burning in their core \citep{Baraffe08,MolliereMordasini12,Bodenheimer13} (see chapter by {\it Baraffe et al.}).


\subsection{The Brown dwarf/planet overlapping mass regime}
\label{sec:BDGP}

An mentioned in section \S\ref{BD_IMF}, there is now
ample evidence for the existence of free floating brown dwarfs with masses of the order of a few Jupiter masses in (low extinction) young clusters and in the field, see e.g. \cite{Caballero07}, with a mass distribution consistent with the extension
of the stellar IMF into the BD regime. The brown dwarf and planet mass domains thus
clearly overlap, arguing against a clear mass separation. The fundamentally different mass distribution of exoplanets detected by radial velocity surveys, with the mass function rising
below $\sim 30\,\mjup$ \citep{Mayor11}, in stark contrast with the BD mass distribution (see \S\ref{BD_IMF}) clearly  suggests two distinct populations, with different origins.

Of particularly noticeable interest at this stage are the transiting objects Hat-P-2b, with a mass of 9 $\mjup$ \citep{Bakos07} and Hat-P-20b, with a mass of 7.2 $\mjup$ \citep{Bakos11}. Both objects are  too dense to be brown dwarfs.
Assuming that the observational error bars on the radius are reliable, and given the age inferred for theses system, the observed mass-radius determinations imply significant enrichment in heavy material, revealing their planetary nature \citep{Leconte09,Leconte11}. This shows that
planets at least 9 times more massive than Jupiter, close to the DB limit, can form according to the core-accretion scenario, possibly from the merging of lower mass planet embryos.
Note that while massive objects like HAT-P-20\,b approach the upper limit of the mass distribution predicted by the core accretion scenario \citep{Mordasini12b}, its large metal enrichment ($M_Z\sim 340\,\mearth$, \cite{Leconte11}) certainly excludes formation by gravitational collapse.

According to the arguments developed in this section and in the previous ones, the present IAU definition, based on a clear-cut mass limit between BD's and planets, is clearly incorrect and confusing and should
be abandonned. We come back to this point below.


\section{\textbf{	CONCLUSION}}

Even though it is probably still premature to reach definitive conclusions about brown dwarf and giant planet formation and we must remain open to all possibilities, the confrontation of the various theories with observational constraints described in this review suggests some reasonably sound
conclusions. The ability of BD's to form in isolation and in wide binaries; the similarity of BD number-density in low (e.g. Taurus) and high-density environments, indicating no significant dependence of BD abundances upon stellar density; the emerging observations of isolated proto-BD's and pre-BD cores; the many observational properties shared by young BD's and young stars; the  observed close similarity between the prestellar/BD core mass function (CMF) and the final stellar/BD IMF; the BD IMF been consistent with the natural extension of the same stellar IMF down to the nearly bottom of the BD domain. All these points
provide evidence that dynamical interactions and dense cluster environments, disk fragmentation or photoionizing radiation are {\it not required for BD formation}. In contrast, all these properties are consistent with BD formation being a natural scaled down version of star formation by the turbulence induced fragmentation of molecular clumps, leading to the formation of pre-stellar and pre-BD cores. It is not excluded, however, that, under some specific circumstances (very dense environment, very massive disks), the other aforementioned mechanisms might play some role but,
in the absence of clear observational evidence for this so far, they seem unlikely to be the dominant
mechanisms for star and BD formation. 

Conversely, the overwhelming majority of planet discoveries are consistent with planet formation by core accretion. As examined in section \ref{sect:formation_GP}, this scenario migh also explain
planet formation at large orbital distances, not mentioning the possibility to explain such objects
by planet scattering or outward migration. 
Here again, alternative scenarios like disk fragmentation might occur in some places, notably in massive circumbinary disks, and thus explain some fraction of the planet population (possibly $\sim$ 10\% or so, e.g. \cite{Vorobyov13}), but they can hardly be considered as dominant scenarios for planet formation. Hybrid scenarios invoking both gravitational fragmentation at large orbital distances followed by inward migration, invoking even in some cases evaporation (see chapter by {\it Helled et al.}) seem to raise even more problems than they bring solutions, as migration can only exacerbate problems with the gravitational instability. Not
only one needs to form planets by GI but one needs to prevent them to migrate rapidly all the way into the star \citep{VorobyovBasu06,Machida11,Baruteau11, Vorobyov13}. Migration, however, is very likely and fast given that, in that case, the disk mass must be very high, a requirement for GI to occur. If planets indeed form this way, this requires very fine tuning conditions, making the branching ratio for this route very small.

Finally, as discussed in \S\ref{def}, there is ample evidence that the planet and brown dwarf domains overlap and that deuterium burning plays no particular role in the formation process. There is now growing evidence, 
possibly including the WISE survey for the field population (see \S2.1), for the existence of non-deuterium burning free floating brown dwarfs and, conversely, no physical arguments against the possibility for genuine planets to ignite D-burning in their core. This again shows that the
IAU definition has no scientific justification and only brings scientific and mediatic confusion. This also argues against
the use of specific appelations for free-floating objects below the D-burning limit, these latter being simply
non D-burning brown dwarfs.

Given the arguments examined along this review, it seems rather secure to argue that BD's and GP's represent two distinct populations of astrophysical bodies which arise {\it dominantly} from two 
different formation mechanisms.
While BD's appear to form preferentially like stars, from the gravoturbulent fragmentation of a parent (possibly filamentary) molecular clump, GP's arguably form essentially by core accretion in a protoplanetary disk, i.e. from the
growth of solids (planetesimals, pebbles) yielding eventually the accretion of a surrounding gas 
rich H/He envelope. So, the very {\it definition} of a brown dwarf or a giant planet is intrinsically, tightly linked to its formation mechanism. 
As briefly discussed below and in \S\ref{sec:DI}, this latter should leave
imprints which might be observationally detectable. 
As examined in \S\ref{hot_cold}, the luminosity or effective temperature at early
 ages, however, cannot be used as a diagnostic to distinguish between these two populations. 
 On the other hand, we argue in this review that planets are necessarily companions of a central, significantly more massive object.
Consequently, free floating objects down to a few $\mjup$ can rather unambiguously be identified as genuine (non D-burning) {\it brown dwarfs}, and one should stop giving them different names, which
simply adds to the confusion.
This is the most direct observational distinction between the two types of objects. Ejected planets probably exist and weaken this statement but they are unlikely to represent a significant
fraction of the population.
The diagnostic is less clear for wide companions to stars, except if the mass ratio can safely exclude one of the two possibilities. According to the distinct formation scenarios, planets should have a substantial enrichment in heavy elements compared with their parent star, as observed for our own solar giant planets, whereas BD's of the same mass should have the same composition as their parent cloud. As  suggested in \cite{Chabrier07} and  \cite{Fortney08} 
 and as examined in \S\ref{sec:DI}, giant planets should bear the signature of this enhanced 
 metallicity in their atmosphere. Spectroscopy or even photometry of atmospheric chemical abundances, notably metal-dominated compounds like e.g. CO or CO$_2$ may provide clues about 
 the formation mechanism and thus help identifying the nature of the object.
In the absence (so far) of 
clear observational diagnostics (mean density, atmospheric abundances, oblateness, etc.) to get clues 
about these formation conditions, the very nature of some
of these objects might remain uncertain. As frustrating as this may sound, we will have to admit such present uncertainties, as ignorance is sometimes part of science.

\textbf{Acknowledgments.} 
The authors are grateful to the referee, A. Morbidelli, for a careful reading of the manuscript and constructive comments.
The research leading to these results has received funding from the European Research Council under the European Community 7th Framework Programme (FP7/2007-2013 Grant Agreement no. 247060).

  
\bibliographystyle{ppvi_lim1.bst}    
\bibliography{ms.bib}         

\end{document}